\documentclass{aa}  
\usepackage{graphicx}
\usepackage{txfonts}
\usepackage{natbib}
\usepackage{amsmath}
\usepackage{graphicx}
\usepackage[colorlinks=true, allcolors=blue]{hyperref}
\usepackage{rotating}
\usepackage{caption}
\def\Feka{Fe~K$\alpha$}
\def\pdfrho{${\rm PDF} (\rho)$}
\def\pdfs{${\rm PDF} (s)$}
\def\sgras{Sgr~A$^\star$}
\def\chandra{{\it Chandra}}
\def\xmm{{\it XMM-Newton}}
\def\nustar{{\it NuSTAR}}
\def\ixpe{{\it IXPE}}

\usepackage{makecell}

\author{G. Stel, G. Ponti, F. Haardt}

\begin{document}

\title{25 years of XMM-Newton observations of the Sgr A complex}

   \subtitle{3D distribution and internal structure of the clouds}

\author{Giovanni Stel\inst{1,2}
        \and Gabriele Ponti\inst{2,3,1} 
        \and Francesco Haardt\inst{1,2,4}
        \and Mattia Sormani\inst{1}
        }

\institute{
        Como Lake Center for Astrophysics (CLAP), DiSAT, Universit\`a degli Studi dell'Insubria, via Valleggio 11, I-22100 Como, Italy \\
              \email{giovanni.stel@inaf.it}
\and
        INAF -- Osservatorio Astronomico di Brera, Via E. Bianchi 46, 23807 Merate, Italy        
\and 
        Max-Planck-Institut für extraterrestrische Physik, Giessenbachstrasse, 85748, Garching, Germany
\and
        INFN, Sezione Milano-Bicocca, P.za della Scienza 3, I-20126 Milano, Italy}

   \date{Received 3 July 2024; accepted 14 January 2025}

\abstract
   {\sgras, the supermassive black hole at the center of the Milky Way, is currently very faint. However, X-ray radiation reflected by the Sgr A complex, a group of nearby molecular clouds, suggests that it went through one or more periods of high activity some hundreds of years ago.}
   {We aim to determine whether previously proposed physical scenarios are consistent with the observed X-ray variability over the past 25 years. Furthermore, we seek to characterize the spatial distribution, shape, and internal structure of the clouds.}
   {We exploit the whole set of available \xmm\ observations of the Sgr A complex to date, extending the previously studied dataset on variability by at least 12 years. Starting from the recent \ixpe\ result that places the so-called Bridge cloud 26 pc behind \sgras, we reconstruct the Line-Of-Sight (LOS) position of the remaining clouds in the molecular complex, assuming that they were illuminated by a single flare. Additionally, we derive the probability density function (PDF) of the molecular density. We also study the 3D geometry of the complex in case two flares illuminate the clouds.}
   {As of spring 2024, the lightfront is still illuminating the Sgr A complex, with the Bridge currently being the brightest cloud. The other clouds in the complex have faded significantly. In the single flare scenario, the Sgr A complex is located $\simeq 25$ pc behind \sgras. In the past 25 years, the illuminated region spans 10-15 pc along the LOS. The derived PDF of the molecular hydrogen exhibits a roughly log-normal distribution, consistent with previous \chandra\ results, with a potential excess at the high-density end.}
   { Both a single and a multiple flares scenario can explain the observed X-ray variability. Previous concerns about the single flare scenario, raised by shorter monitoring, are now overcome in the 25 years of monitoring. If two flares illuminate the clouds, they must be separated by at least $\sim 30$ years. We speculate that these clouds are closer to \sgras\ than the nuclear molecular ring at $\simeq 100 - 200$ pc  and possibly drifting from the ring to the inner region of the Galaxy.}


   \maketitle

\section{Introduction}
The highest concentration of giant molecular clouds in the Milky Way is found in its inner few hundred parsecs, in the so-called Central Molecular Zone \citep[CMZ,][]{Morris1996}. The CMZ is characterized by  unique physical conditions within the Galaxy, with a high average gas density ($\sim 10^4$ cm$^{-3}$), a typical molecular gas temperature of 50-100 K \citep{Mills2017}, and a magnetic field strength of order $\sim 1$ mG in the densest clouds \citep{Ferriere2009,Mangilli2019,Tress2024}. Molecular clouds are arranged in a ring-like structure located 100-200 pc from the center \citep{Sofue1995,Molinari2011,Kruijssen2015}, similarly as observed around the center of other barred galaxies \citep{Comeron2010,Stuber2023}. This ring-like structure forms by the accumulation of molecular gas at the inner edge of an extended gap around the inner Lindblad resonance of the Milky Way's bar \citep{Sormani2024}, i.e., the radius at which the epicyclic frequency of radial oscillations is twice the forcing frequency of the bar seen by a particle in the bar's rotating frame.

The advent of X-ray astronomy has shown that some 
molecular clouds in the CMZ emit in the X-ray band \citep{Sunyaev1993, Markevitch1993}. The emission spectrum is characterized by a scattered continuum, and a bright, fluorescent \Feka\ line at 6.4 keV \citep{Koyama1996, Sidoli2001}. Interaction with cosmic rays and/or with radiation from past outbursts have been proposed as the physical mechanisms that illuminate these clouds \citep{Sunyaev1998, Yusef-Zadeh2007}. The X-ray monitoring of the CMZ has shown that variability of the \Feka\ line occurs on short time scales ($\sim 1$ year), thus suggesting that at least a considerable fraction of the line emission is due to reprocessed radiation from a past energetic outburst of comparable duration \citep{Muno2007, Ponti2010, Ponti2013}.

The primary candidate as the illuminating source is \sgras\ itself. The supermassive black hole (SMBH) at the center of the Milky Way \citep{Schodel2002, Ghez2003, Genzel2003, Do2019, EHT2022, Gravity2023} is currently faint, with an X-ray luminosity of $L_X\sim 10^{33}$ erg s$^{-1}$ \citep{Baganoff2003}, a bolometric luminosity of $L_{\rm bol} \sim 10^{36}$ erg s$^{-1}$ (corresponding to $\sim 10^{-9}$ its Eddington luminosity). \sgras\ shows daily X-ray flares \citep{Baganoff2001, Neilsen2013, Ponti2015, Mossoux2020}, during which its luminosity can increase by more than a factor of 100 \citep{Porquet2003, Porquet2008, Nowak2012, Haggard2019}. The low luminosity and short durations (about 1 hour) of these flares do not generate a detectable signal from the molecular clouds in the Galactic center. However, despite its current low average luminosity, the intensity of the reflected light indicates that  \sgras\ reached a luminosity of at least $L_X\sim 10^{39}$ erg s$^{-1}$  in the past hundreds of years \citep{Sunyaev1998, Ponti2010, Churazov2017a, Marin2023}. The observed fluorescence is then the {\it echo radiation} of such an energetic event.

Sgr B2 is the most massive molecular cloud in the Galactic center and the first studied in the X-ray band since the '90s \citep{Koyama1996}. Its X-ray non-thermal emission has been fading since 2001 \citep{Terrier2010, Zhang2015, Kuznetsova2022}, reaching an almost constant value. Fluorescence variability was also observed in Sgr C, confirming its reflection origin \citep{Chuard2018}. Similarly, the Sgr A complex, the molecular clouds lying within $\simeq 15$ arcmin to \sgras\ (projected distance of 35 pc at 8.2 kpc), was widely studied \citep{Muno2004, Park2004, Khabibullin2022}.

Comprehensive studies of the variability of individual clouds within the Sgr A complex have been conducted using both Chandra \citep{Muno2007, Clavel2013}, utilizing data up to 2011, and \xmm\ \citep{Ponti2010, Capelli2012, Terrier2018}, with data extending through 2012. \xmm\ is uniquely suited for studying the X-ray variability of the Sgr A complex, as this telescope has observed the Galactic center, including the Sgr A complex, nearly every year. In this work, we extend the analysis of variability to the present (Spring 2024), incorporating approximately 12 additional years of observations, thereby doubling the monitored period.

Currently, the Sgr A complex is the brightest sample of molecular clouds emitting in the X-ray band \citep{Terrier2018, Khabibullin2022}. High-resolution \chandra\ observations of this region showed that in single clumps, the \Feka\ emission can light up and be extinguished in about two years \citep{Clavel2013, Churazov2017a}, suggesting this as an upper limit for the duration of at least one flare that is illuminating the clouds. \citet{Clavel2013} showed that the echo radiation from other molecular clouds in the region evolves on longer time scales (8-10 years). They also observed that the densest clouds in the complex featured an \Feka\ emission comparable with other less dense clouds. Such behavior can not be explained by individual clouds illuminated by a single flare, since the densest knots should have the highest \Feka\ intensity. Thus,  \citet{Clavel2013} proposed a {\it multiple flares scenario} characterized by two flares of different duration. In this picture, depending on their distance from \sgras, different clouds are illuminated by one of at least two different past flares. Nevertheless, the different time scales on which the fluorescence evolves can also be related to the internal structure of the molecular complex \citep{Churazov2017a}. Indeed, if the gas distribution is clumpy, with high-density cores surrounded by low-density envelopes, the variability is expected to occur on two different time scales, even in the case of a single short flare. In fact, the signal in the smaller cores should evolve on shorter time scales compared to the more diffuse envelopes.

Recently, an observation of a subset of the Sgr A complex (the so-called Bridge cloud) was performed by the {\it Imaging X-ray Polarimetry Explorer} \citep[\ixpe,][]{Weisskopf2022}, showing that the continuum-reflected light is linearly polarized, as theoretically predicted \citep{Churazov2002,Marin2014,Marin2015,Churazov2017b}. This measurement, along with the short time variability observed in the past years, definitely proves the reflection origin of the observed radiation. Moreover, the measured polarization angle ($-48^\circ \pm 11^\circ$\footnote{Polarization angle is taken anti-clockwise from North in the equatorial coordinate system.}) is consistent with \sgras\ being the illuminating source. The polarization degree implies a position of the illuminated Bridge cloud about 26 pc behind \sgras\ along the LOS. This is equivalent to an estimated age of the flare of $205^{+50}_{-30}$ years \citep{Marin2023}. 

The intensity of the reflected light depends on the illuminating source flux, the cloud properties (i.e., density, size, and iron abundance), and the relative positions of the source and the clouds \citep{Sunyaev1998}. It is possible to reconstruct the 3D cloud distribution {\it inside} the molecular complex by matching the time and space variability of
the signal over the years \citep{Churazov2017a}. Since the duration of the illuminating flare is expected to be much shorter than the typical light crossing time of the Sgr A complex, the wavefront is practically scanning the cloud distribution, revealing the internal structure of the molecular complex. The X-ray intensity is directly related to the molecular gas density. Hence the X-ray
monitoring of Sgr A provides insights into the gas density distribution
 \emph{inside} the molecular complex \citep{Churazov2017c}.
The molecular gas density distribution is particularly interesting since it is a key ingredient of star formation models \citep{Hennebelle+2012}. In this context, the CMZ represents a peculiar environment in which, despite the large concentration of molecular material, star formation is less effective than in other parts of the Milky Way or other galaxies \citep{Longmore2013,Henshaw2023}.

In addition to the variable X-ray emission, molecular clouds could also shine with an almost constant brightness. This residual emission could arise from the interaction of the gas with cosmic rays \citep{Yusef2007, Dogiel2009, Tatischeff2012}, or from multiple scattering of photons inside the clouds \citep{Sunyaev1998, Molaro2016}. 

The present work focuses on the Sgr A complex, the ensemble of clouds located Northeast of \sgras\, within 15 arcmin from the SMBH. We aim to verify if previous scenarios based on 10-12 years of monitoring can still describe the observed X-ray variability in the longer 25-year period we consider. We describe the dataset and observed variability in Sect.~\ref{sec:data_reduction} and \ref{sec:variability}. We first study the single flare scenario. The key assumption of our analysis is that the Sgr A complex is illuminated by an energetic event that occurred some 200 years ago, as measured by \ixpe. We derive the implications of the observed variability, and we reconstruct the 3D spatial distribution in Sect.~\ref{sec:geometry}. We derive the  PDF of the density of the molecular gas in Sect.~\ref{sec:pdf}. In Sect.~\ref{sec:multiflare}, we discuss and compare the multiple flares scenario with the  single flare scenario. In Sect.~\ref{sec:20-50}, we briefly discuss the constraints on past \sgras\ flaring activity from the closest massive molecular clouds.

\section{Data reduction and imaging}
\label{sec:data_reduction}

We analyzed the complete set of XMM-Newton observations, which included \sgras\ in their Field-Of-View (FOV) from 2000 to 2024. Raw data are reprocessed using the standard \xmm\ pipelines in the Scientific Analysis System (SAS 21.0.0), along with the last release (as of April 2024) of the current calibration files (CCF). As we are mainly interested in the emission of the \Feka\ line, we created exposure-corrected images by selecting those events whose energy is comprised between $E_1=6.1$\,keV and $E_2=6.6$\,keV, with the expression \texttt{"\#XMMEA\_EP $\& \&$ (PATTERN<=4)"}. The energy range is chosen to take into account the spectral resolution of the instrument, which is about 120 eV.
Images were corrected for out-of-time events. We merged images belonging to the same year, obtaining a total of 16 maps. In order to have sufficient homogeneous statistics through the years, we combined the observations from 2000 and 2001. Regions with exposure lower than 25\% of the maximum value are masked (gaps between CCD, usually 1-3 pixels depending on the position on the detector, were masked too). Images were then smoothed with a Gaussian kernel ($\sigma = 2.5$ pixels), using the \texttt{astropy.convolution} module\footnote{\url{www.astropy.org} and \url{https://docs.astropy.org/en/stable/convolution/index.html}}  which allows the reconstruction of the space between sub-arrays of the CCD by kernel-based interpolation. The width of the Gaussian kernel is comparable to the PSF of the instrument. 

\begin{figure}
\centering
\includegraphics[width=\linewidth]{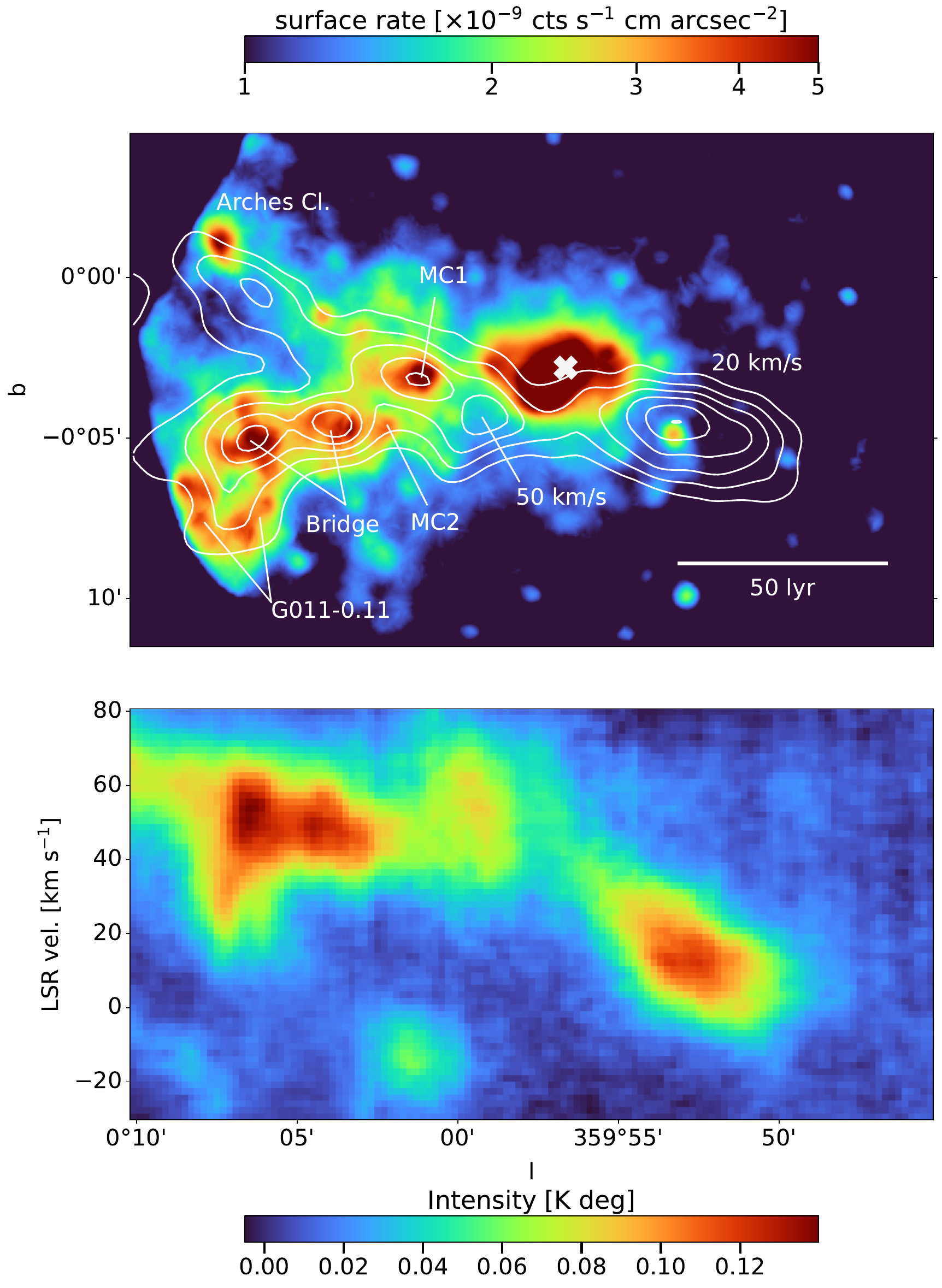}
\caption{{\it Top panel:} Median of the Sgr A complex emission in the \Feka\ 6.4 keV line, in the period 2000 - 2024. The white cross marks the position of \sgras. Contours are computed from the N$_2$H$^+$ line intensity, obtained with the MOPRA telescope, integrated over [-30; 80] km s$^{-1}$ range of velocities \citep{Jones2012}. Contour lines refer to 20, 30, 40, 50, 60, and 70 K km s$^{-1}$. The central strong emission, encompassing \sgras, not associated with molecular clouds, is the supernova remnant Sgr A east \citep{Maeda2002}. {\it Bottom panel:} MOPRA longitude-velocity diagram of the same region above. The latitude has been integrated over the [-0.193; 0.169] deg range. } 
\label{fig:median}
\end{figure}

Fig.~\ref{fig:median} (top panel) displays the median of all the images\footnote{ Throughout the article, we show the images obtained with the EPIC-pn camera only. EPIC-mos data are used in the analysis to verify that the results are not affected by systematics in the map construction.}. The diffuse emission due to the reflected light is clearly visible toward positive longitudes. In the same figure, we overplotted the N$_2$H$^+$ line intensity contours \citep{Jones2012}, which tracks the molecular clouds column density. We labeled the names of individual clouds forming the Sgr A complex; the nomenclature follows the one adopted in previous X-rays-based works \citep{Ponti2013, Clavel2013}. The bottom panel of the same figure shows the longitude-velocity diagram of the same region, in which the individual clouds forming the molecular complex can be identified. From the right, the {\it 20 km/s molecular cloud} can be appreciated; farther east, there is the {\it 50 km/s cloud} peak. Toward positive longitudes, two clumps characterize the so-called {\it Bridge} structure, with velocity in the range of $40 - 60$ km s$^{-1}$. {\it G0.11-0.11}, the cloud, which is V-shaped in the X-ray band, is characterized by lower velocities. The last two clouds collapsed in a single blob, in the longitude-velocity diagram, because of the integration over the range of latitudes the Sgr A complex spans. In the same diagram, {\it MC1} appears separated from the other molecular clouds, lying between -20 km s$^{-1}$ and 0 km s$^{-1}$. A similar velocity characterizes the {\it MC2} cloud, which appears in the X-ray but is not clearly identified in the radio contours. Toward positive Galactic latitudes, at the FOV limit of the observations, a bright excess identifies the position of {\it Arches Cluster} molecular cloud. It is named after the more famous dense star cluster, which occupies a similar projected direction in the sky. The cloud's velocity is about $-25$ km s$^{-1}$.

The observed X-ray emission is mainly due to two components, i.e., the reflection from molecular clouds superimposed to a background glare of point and extended sources \citep{Ponti2013, Ponti2015, Koyama2018, Anastasopoulou2023}. Additionally, the reflection component at $\simeq 6.4$\,keV is due to the \Feka\ line and the scattered continuum underneath. Since we are interested in the flux originating from the line, we proceeded to isolate such a signal. The continuum spectrum underneath the line can be approximated by a power law with the same slope as the illuminating source. Assuming that the effective area is constant in $[E_1;E_2]$, then the continuum photon flux, $\mathcal{C}$, in the same energy band is related to the \Feka\ photon flux ($\mathcal{L}$) by:

\begin{equation}
\label{eq:line_continuum}
    \mathcal{C} = \frac{E_{K\alpha}^\Gamma (E_1^{1-\Gamma} - E_2^{1-\Gamma})}{(\Gamma-1)W_{K\alpha}}\,\times \, \mathcal{L},  
\end{equation}
where $E_{K\alpha}$ and $W_{K\alpha}$ are the energy and the equivalent width of the \Feka\ line, respectively, and $\Gamma$ is the photon index of the continuum spectrum.
The equivalent width of the \Feka\ line is predicted and measured to be $W_{K\alpha}\approx 1$ keV \citep{Sunyaev1998, Ponti2010}. Assuming $\Gamma=2$ \citep{Ponti2010}, $\mathcal{C} \simeq 0.5 \times \mathcal{L}$. Therefore, the \Feka\ flux represents roughly 2/3 of the reflection flux between 6.1 and 6.6 keV.

In order to remove the background, we estimated its value by a simple extrapolation in the [6.1;6.6]\,keV band of the intensity measured in the [4.5;6]\,keV band. Note that we avoided the [6.0;6.1]\, keV data as  background and line counts are difficult to disentangle in this range,. In extrapolating the background from lower energies, we employed the Xspec software \citep{Arnaud1996}, assuming a collisionally-ionized diffuse gas (\texttt{apec} model\footnote{\url{http://atomdb.org/}}) with a $kT=8$\,keV temperature and solar abundances \citep[using abundances from][]{Wilms2000}.

\begin{figure*}
\centering
\includegraphics[width=1.0\textwidth]{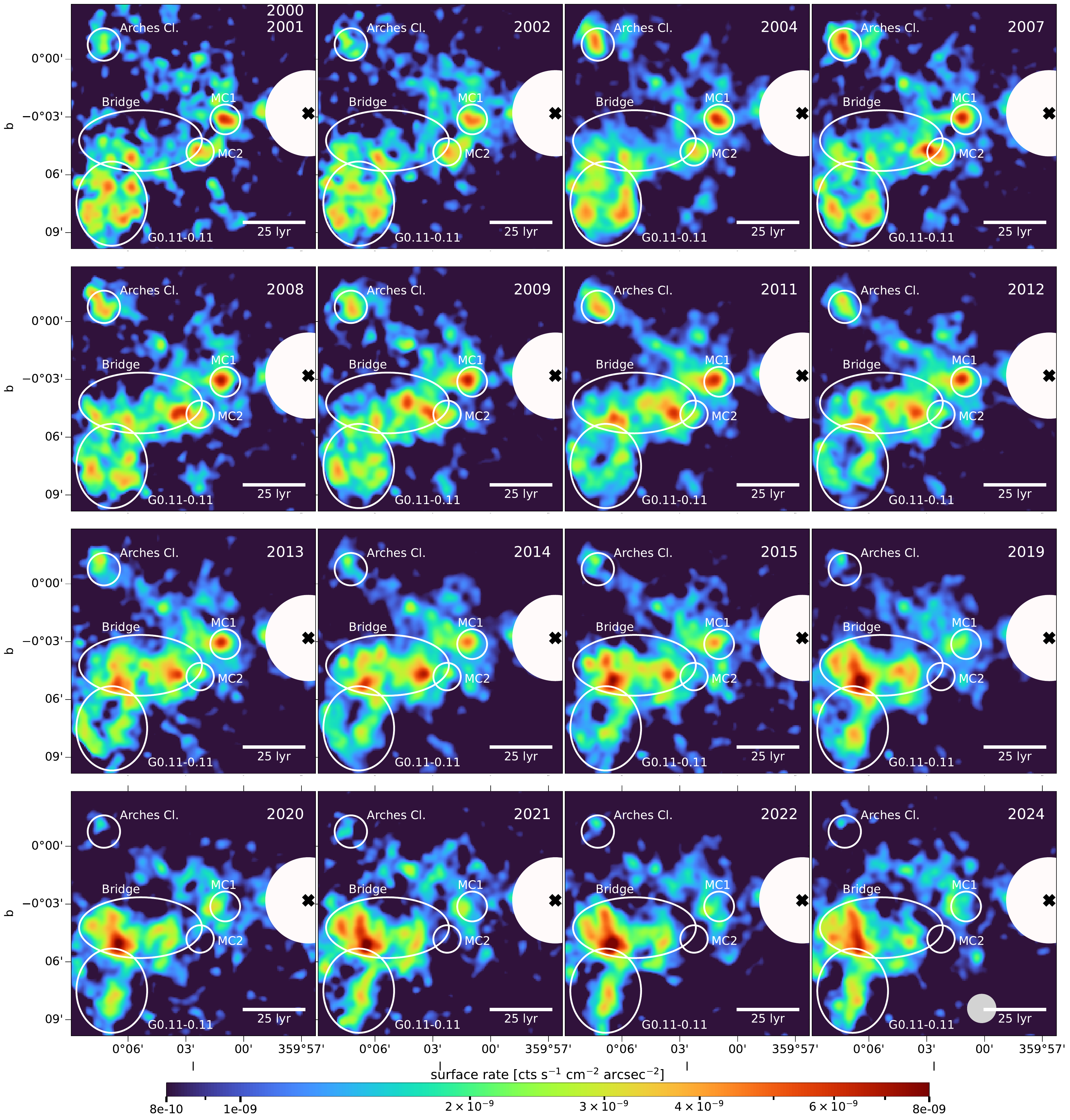}
\caption{\Feka\ emission for different years. The images are created by selecting events in the $[6.1;6.6]$ keV energy range. For each image, a background is computed in the $[4.5;6.0]$ keV band and subtracted. The black cross marks the position of \sgras\, while a recently observed low-luminosity transient \citep{Reynolds2024} is masked during 2024 observation (gray circle). The position of the main molecular clouds is highlighted by ellipses.} 
\label{fig:variability_kalpha}
\end{figure*}

The resulting \Feka\ maps are shown in Fig.~\ref{fig:variability_kalpha}. In the maps, we highlighted the position of the molecular clouds studied in previous works. A black cross marks the position of \sgras, the inner region is masked for the presence of transients. A gray circular region masks the position of an active source, Swift J174610-290018, first detected in February 2024 \citep{Reynolds2024} by Neil-Gehrels Swift/XRT, during the daily monitoring of the Galactic center performed by this telescope \citep{Degenaar2015}. The source is a low-luminosity transient, with an initial X-ray luminosity of $L_X \simeq 4.7\times 10^{34}$ erg s $^{-1}$, at a fiducial distance of 8 kpc; its Equatorial (J2000) coordinates are (17:46:10.4, -29:00:17.6).

\begin{figure*}
\centering
\includegraphics[width=1.0\linewidth]{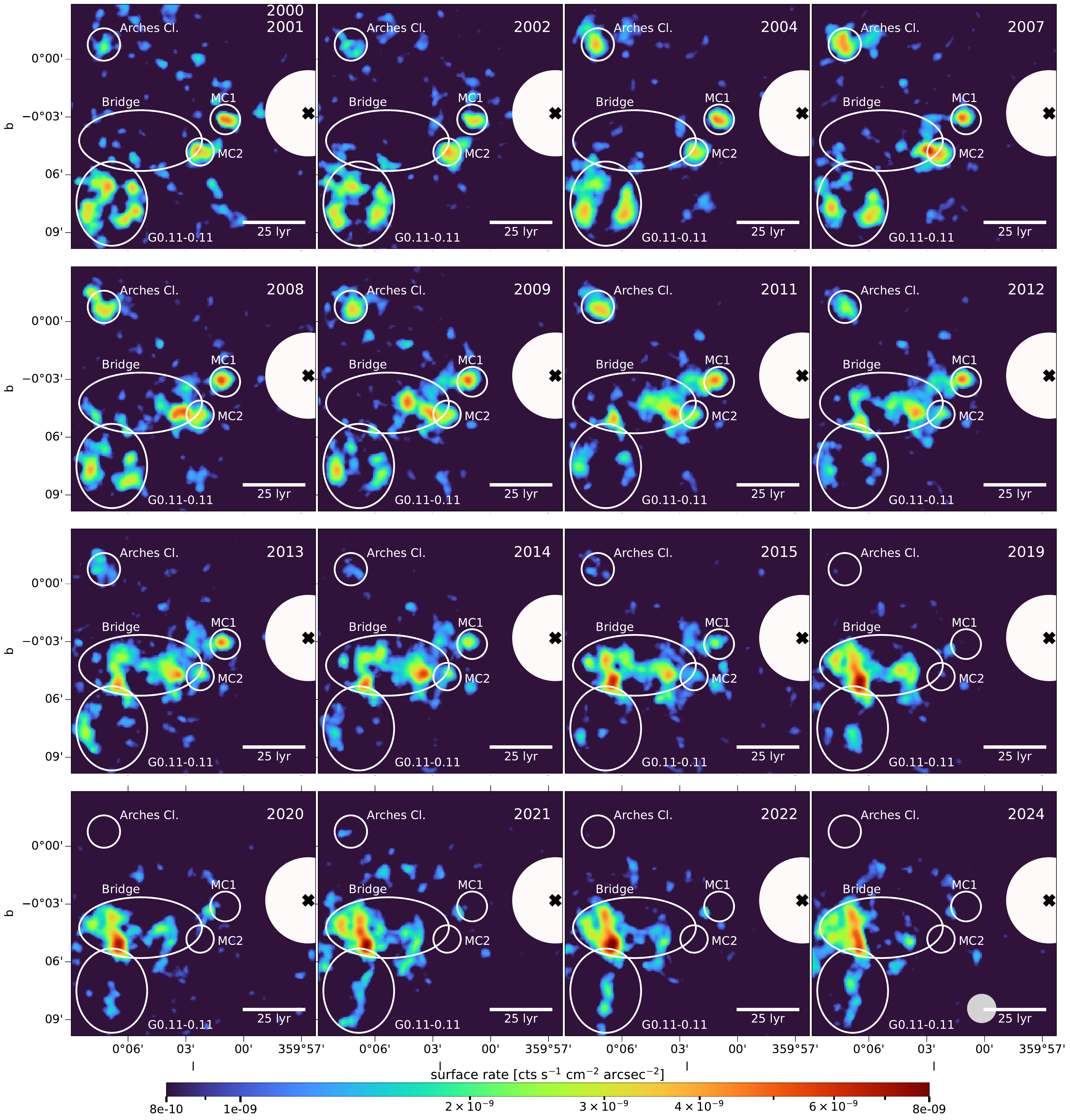}
\caption{Same maps as in Fig.~\ref{fig:variability_kalpha}, but a common minimum map has been subtracted to emphasize the \Feka\ variability. The minimum map is displayed in Fig.~\ref{fig:min}.} 
\label{fig:variability_kalpha_sub}
\end{figure*}

\begin{figure}
    \centering
    \includegraphics[width=1.0\linewidth]{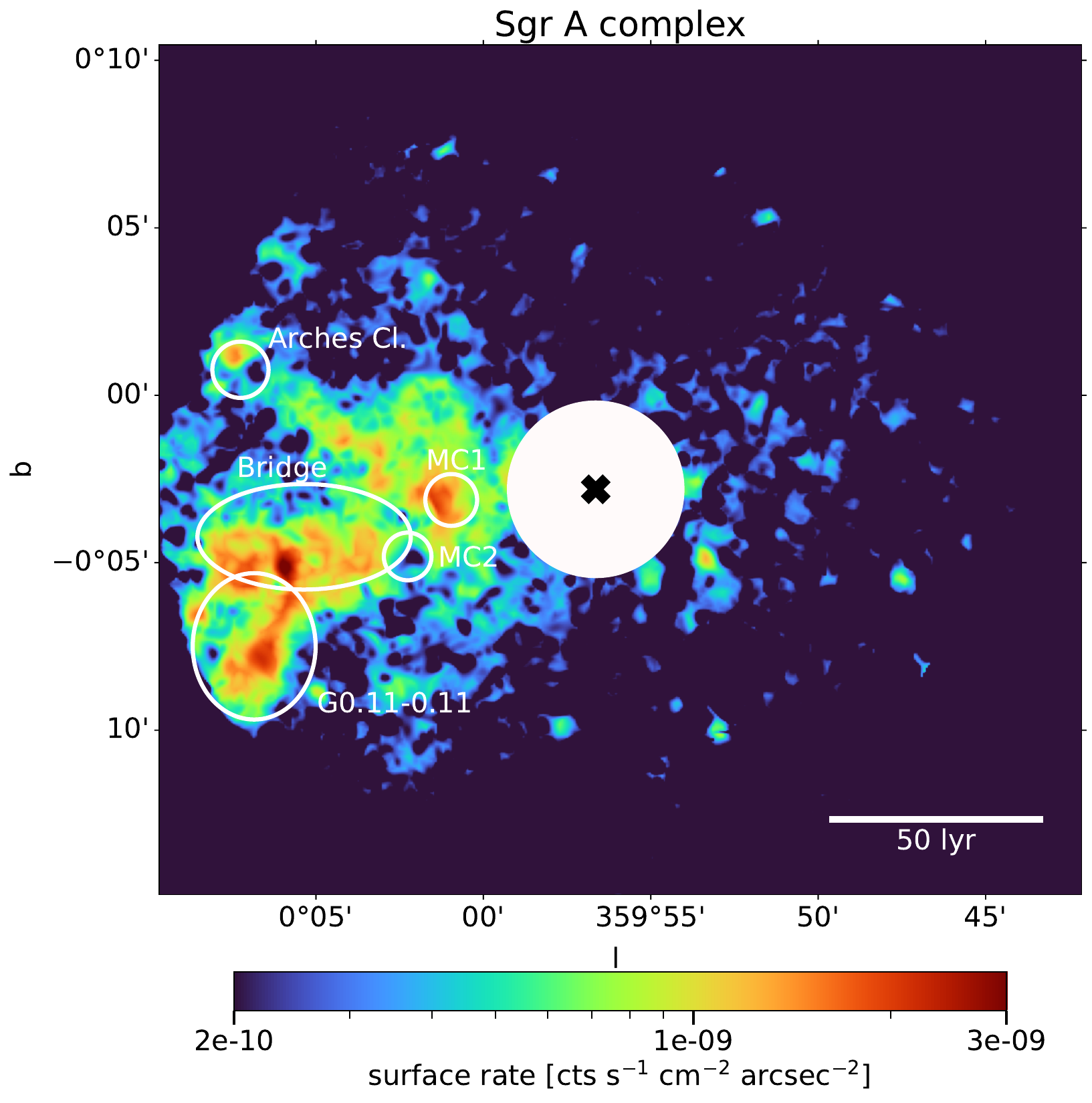}
    \caption{Minimum map of the \Feka\ emission in the period 2000-2024. The map is computed from the maps in Fig.\ref{fig:variability_kalpha}; the value in each pixel is the minimum across the 16 epochs considered in this work. The scale bar indicating the 50 lyr measure across the sky is computed at the Galactic center distance.}
    \label{fig:min}
\end{figure}

Fig.~\ref{fig:variability_kalpha} clearly shows that in some regions, the \Feka\ intensity is different even between consecutive years, whereas other regions present an almost constant emission value. As mentioned before, this almost {\it constant} emission can be due to cosmic rays, or arise in a multiple flares scenario with a second flare of longer duration, or be present if compact cores are surrounded by less dense envelopes. In order to highlight the variability of the \Feka\ emission, we subtracted an estimate of the non-variable component. To do so, we defined a "common reference minimum map" by considering, in each pixel, the minimum across all epochs of the \Feka\ line emission. Fig.~\ref{fig:variability_kalpha_sub} shows the resulting maps. Individual clumps are now more evident. Fig.~\ref{fig:min} displays the aforementioned reference minimum map common to all the epochs. Overall, the Sgr A region is not consistent with a null flux, and variability is occurring up to 2024 (Fig.~\ref{fig:variability_kalpha_sub}). This fact indicates that the radiation of the past flare is still illuminating the Sgr A complex. Part of the emission present in the minimum map could also be associated with a constant component. The surface rate of the minimum map ranges between $\simeq 5 \times 10^{-10}$ and $\simeq 3 \times 10^{-9}$ cts s$^{-1}$ cm$^{-2}$ arcsec$^{-2}$, depending on the region inside the Sgr A complex. The minimum map provides an upper limit of the potential constant component.

\begin{figure}
    \centering
    \includegraphics[width=\linewidth]{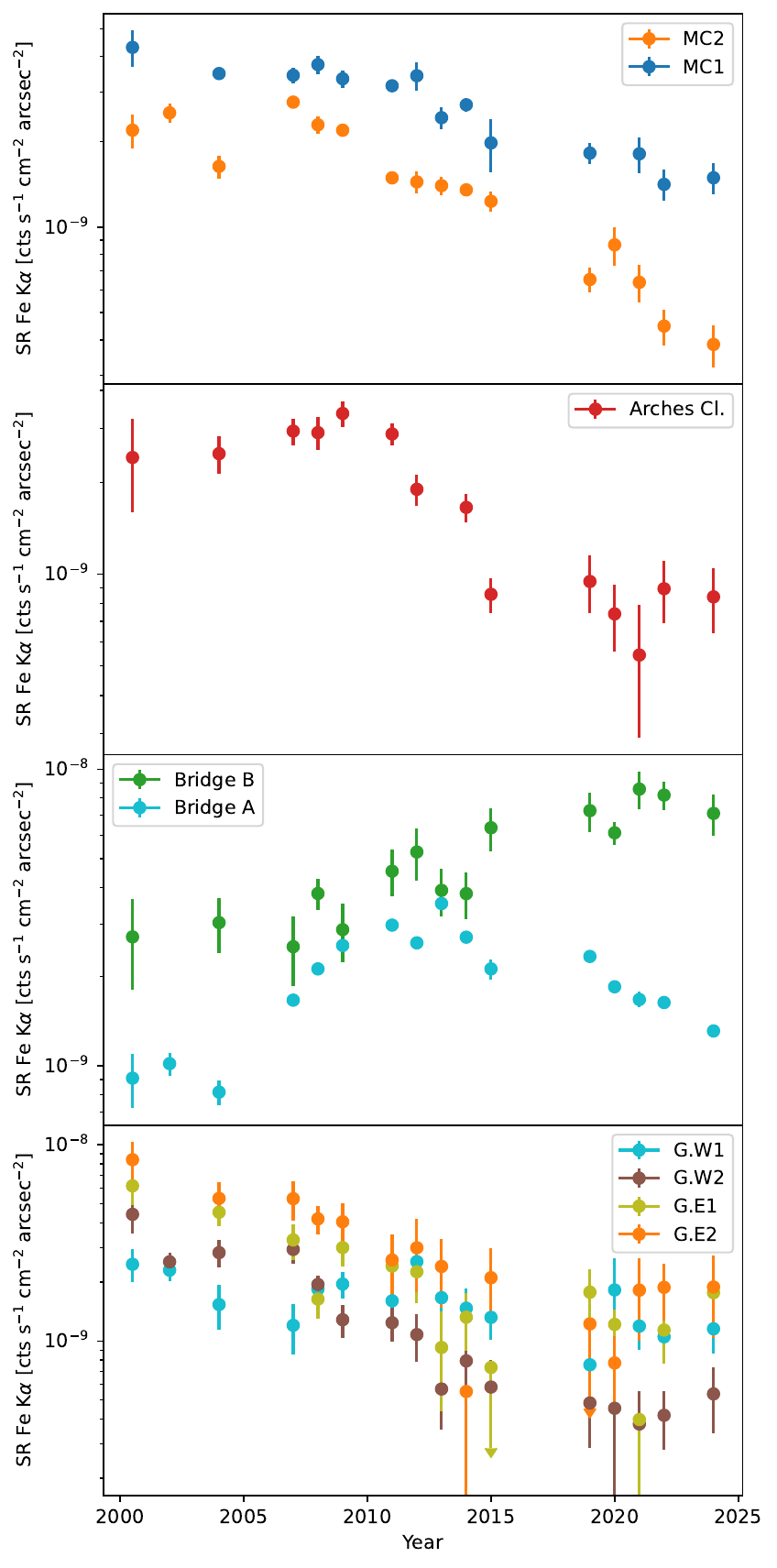}
    \caption{\Feka\ surface rate (the flux is divided by the solid angle over which the spectrum is accumulated) obtained from spectral fitting of selected regions inside the Sgr A complex as a function of the year of observation. Error bars refer to 1 $\sigma$ uncertainty.}
    \label{fig:spectral_variability}
\end{figure}

In addition to the presented maps, we computed the spectra for some selected regions of interest. The regions, and the evolution of the \Feka\ intensity derived from the spectral fit, are described in the next Section, and displayed in Fig.~\ref{fig:spectral_variability}. The regions have been chosen to highlight with light curves the description of the most interesting clouds as derived by the images. However, a fixed region to describe a variable-shaped emission can only partially characterize the \Feka\ evolution. For this reason, the presented set of 16 maps is the more accurate source to describe the fluorescence variability. For each observation, the spectrum is computed for one of the three CCDs on board of \xmm, depending on which instrument is better suited to accumulate the spectrum of that specific region (we checked if the region is entirely within the CCD, and it is not significantly affected by defects of the CCD such as bad pixels or dark columns). Spectra are computed for the observations with a filtered exposure greater than 15 ks. 

The spectral fit is performed with Xspec in two steps. First, we fitted the whole set of observations of a specific region, in the 4.5-7.8 keV band, with a phenomenological model composed of a thermal component and a reflection component. Specifically, in Xspec the model is \texttt{tbabs*(apec+tbabs*powerlaw + gauss)}. The temperature of the \texttt{apec} component is fixed at 8.0 keV, according to the value used to reconstruct the set of 16 maps. The power law is fixed to $\Gamma=2$, the \Feka\ line energy is fixed at 6.4 keV and its width at 0.01 keV, coherently with the values reported in the literature \citep{Ponti2010}. The normalization of the power law and the intensity of the line are left free for each observation, while all the other parameters are commonly fitted to all the observations. Once the background component and the internal absorption of the cloud are determined, we performed a second fit. In this case, we simultaneously fitted all the observations belonging to each of the 16 epochs. To verify that the particular choice of the astrophysical background does not affect the flux of the \Feka\ line,  we repeated the spectral fit, assuming a different model for the background. We adopted the same model described in \citet{Chuard2018}, where the background is modeled as the sum of two thermal components at fixed temperatures of 1.0 and 6.5 keV. We performed the fit in the 2.0-7.8 keV energy band. The normalization of the \Feka\ is consistent for all the regions across all the epochs within one standard deviation. In Fig.~\ref{fig:spectral_variability}, we reported the surface rate, i.e. the flux of the \Feka\ line divided by the solid angle over which the spectrum is integrated, for the second model of the background, along with the 1 $\sigma$ uncertainty.

\section{Temporal and spatial variability}
\label{sec:variability}

Our work mainly focuses on the five clouds highlighted in Fig.~\ref{fig:variability_kalpha} and Fig.~\ref{fig:variability_kalpha_sub}, labeled as MC1, MC2, the Bridge, the Arches Cluster, and G0.11-0.11.
The position of the X-ray clumps matches the location of the clouds as detected in the radio band.
The several known molecular clouds in the Sgr A complex are discussed below.

\subsection{MC1}
Fig.~\ref{fig:variability_kalpha} and \ref{fig:variability_kalpha_sub} show that MC1 was already illuminated in 2000. Therefore, we do not know when the light front started to illuminate this cloud.
Within the monitored time frame, flux variations of MC1 appear to be slow, i.e. both in the set of images and in the lightcurves, peaks of 1-2 years are not observed. This is partially due to the spatial resolution of the maps. Overall, the \Feka\ surface brightness is almost constant ($\simeq 4 \times 10^{-9}$ cts s$^{-1}$ cm$^{-2}$ arcsec$^{-2}$) until 2014, while it appears to dim later on, reaching a surface rate of $\lesssim 2 \times 10^{-9}$ cts s$^{-1}$ cm$^{-2}$ arcsec$^{-2}$. This behavior is confirmed in the light curve (Fig.~\ref{fig:spectral_variability}). Moreover, the reflection component in both images tends to move toward larger Galactic longitudes, away from \sgras. This behavior is consistent with \sgras\ being the illuminating source.
Our last observation, performed in 2024, still shows some \Feka\ emission arising from MC1. Such behavior is confirmed in the minimum maps (Fig.~\ref{fig:min}), where the intensity in the position of MC1 is clearly not consistent with a null flux.

\subsection{MC2}

MC2 is a molecular cloud with a velocity ranging between -10 and 5 km s$^{-1}$, which is significantly different from the velocities characterizing the bulk of the Bridge cloud ($\simeq 40 - 60$ km s$^{-1}$, see next Sect.~3.3). 
Similarly to MC1, MC2 has been observed to be bright since 2000; it 
is not known when the wavefront started to illuminate the cloud. Similarly to MC1, the brightness of this cloud diminishes over 25 years (Fig.~\ref{fig:variability_kalpha} and \ref{fig:spectral_variability}).  The surface rate drops from  $\simeq 5 \times 10^{-9}$ cts s$^{-1}$ cm$^{-2}$ arcsec$^{-2}$ peaks to less than $\simeq 1 \times 10^{-9}$ cts s$^{-1}$ cm$^{-2}$ arcsec$^{-2}$ in 2024. The minimum map shows some residual emission. 
The important drop of the fluorescence, by a factor 5, indicates that the wavefront of the energetic event that has illuminated this cloud has almost completely passed it.

\subsection{Bridge}
\label{sub:Bridge}
\begin{figure*}
    \centering
    \includegraphics[width=1.1\textwidth]{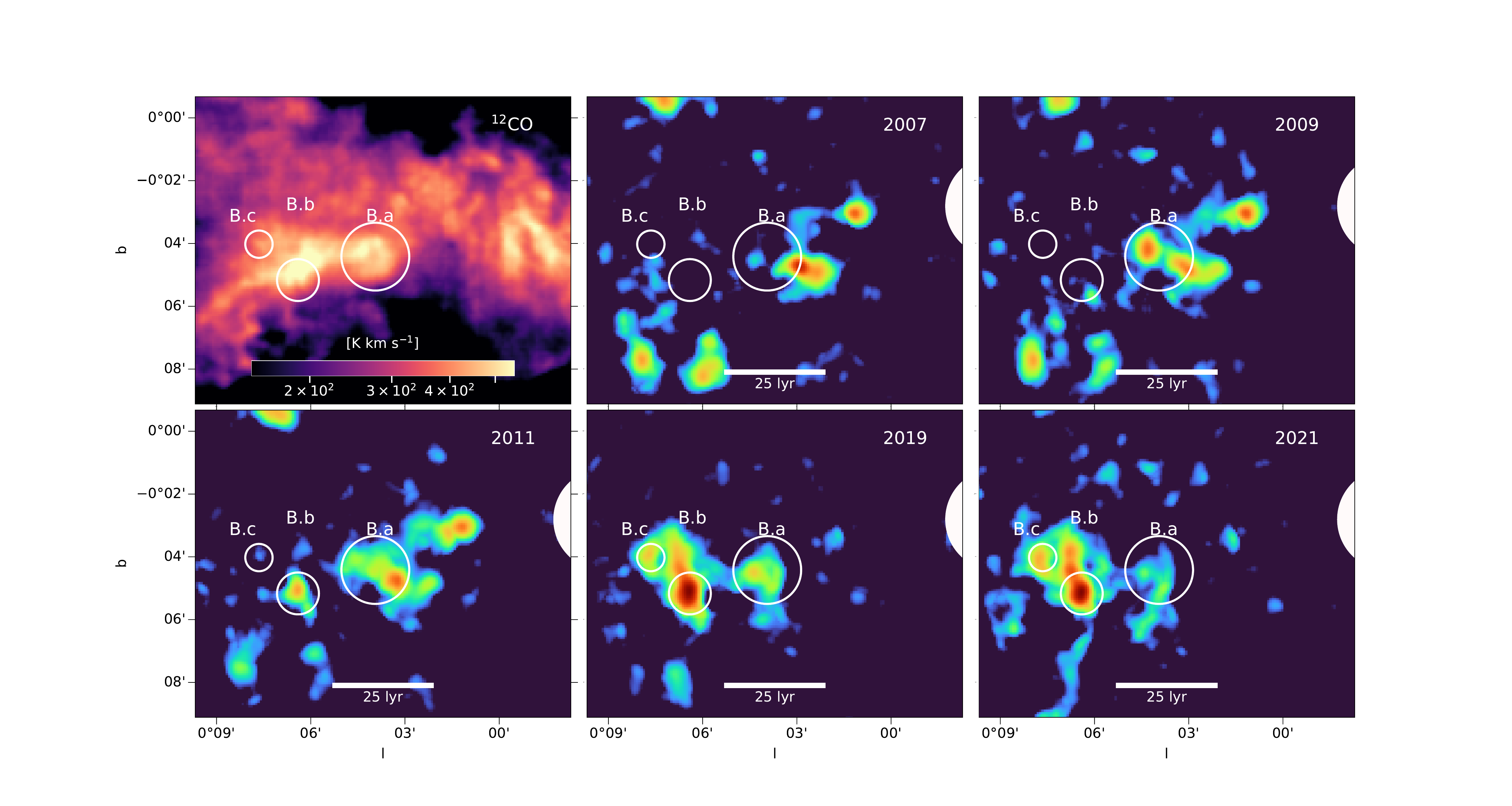}
    \caption{Fluorescence evolution in the Bridge. The upper left image is the $^{12}$CO line intensity in the 45-65 km s$^{-1}$ range \citep{Eden2020}. X-ray colorbar is consistent with Fig.~\ref{fig:variability_kalpha} and \ref{fig:variability_kalpha_sub}.}
    \label{fig:bridge}
\end{figure*}

Details of the evolution of the Bridge structure, with a selection of five particular different snapshots, are shown in Fig.~\ref{fig:bridge}. The upper left panel shows the $^{12}$CO intensity in the 45-65 km s$^{-1}$ range, obtained with the James Clerk Maxwell Telescope \citep{Eden2020}, while the remaining panels show the variable component (\Feka\ emission after the minimum subtraction). The cloud is also known as the "three little pigs" \citep{Battersby2020}, since three different clumps are visible in dust emission, (G0.05–0.07 "Sticks", G0.10–0.08 "Stones", and G0.15–0.09 "Straw"). The first two are also clearly visible in the CO map and in the X-ray ones (B.a and B.b).

As initially reported by \citet{Ponti2010}, a significant variation was observed in the Bridge starting from its westernmost side in 2007 (The "Sticks", B.a in Fig.~\ref{fig:bridge}), close to MC2. The same behavior is confirmed in the light curve (Fig.~\ref{fig:spectral_variability}, panel 3), where a surface rate $\lesssim 10^{-9}$ cts s$^{-1}$ cm$^{-2}$ arcsec$^{-2}$ is measured before 2007, to increase in the subsequent years. In 2008, the emission appeared to be propagating eastwards, it continued to do so over the following years, as expected once an energetic wavefront originating at the \sgras\ position reached the cloud complex. In 2009, two distinct clumps with similar surface rates are visible inside B.a. By looking at these two clumps, \citet{Ponti2010} discovered the superluminal propagation of the \Feka\ emission across the sky. Indeed, the two clumps are separated by a projected distance of about 15 light years, but they become bright just 2 years apart. The apparent superluminal motion of an illuminating front in a molecular cloud is a strong proof that an external source illuminates the clouds. 
Since 2011, a new \Feka\ clump appeared even farther east, in the region labeled as B.b (Fig.~\ref{fig:bridge}). Again, the projected propagation of the signal is superluminal, being these clumps about 20 light years from the B.a region. The region appears to become constantly brighter, while, in the following years, it expanded further eastward, reaching B.c (about 15 light years from B.b) and steadily becoming brighter, reaching a surface rate of $\sim 10^{-8}$ cts s$^{-1}$ cm$^{-2}$ arcsec$^{-2}$. 
The observed superluminal propagation of the signal reported in \citet{Ponti2010} using 2009 \xmm\ data is now confirmed in subsequent years. This fact strongly proves that the variability is induced by an external source. From 2015 an extended tail, almost perpendicular to the plane of the Galaxy, starts to develop in region B.b. Since 2014, the Bridge has appeared as the brightest region of the Sgr A molecular complex in the \Feka\ band.
Note that the evolution of the signal toward positive longitudes is not complete since observations from 2016 to 2018 are missing. 

Beyond superluminal propagation of the \Feka\ emission, the Bridge shows another fundamental feature. In fact, it is the only analyzed cloud in which a clear increase of the \Feka\ emission is observed after the start of the \xmm\ observations. However, since the minimum map (Fig.~\ref{fig:min}) shows clearly a \Feka\ emission in the Bridge position, the region was already illuminated before, and remained illuminated for the entire 25 years of monitoring. Therefore, the X-ray variability in the Bridge is overimposed to a constant (or slowly variant) \Feka\ emission. As mentioned in the Introduction, this second component can be explained by different effects. It can be associated with a less dense envelope illuminated by the same flare, which generates substantial variability in the two main knots. Since the \Feka\ intensity increases by about one order of magnitude in B.b, the envelope may be about one order of magnitude less dense than this knot. Note however, that we are limited by the resolution of the instrument, denser knots can be for example identified with dedicated \chandra\ observations. It is difficult to associate the observed variability with a two flares paradigm because the two flares, with different durations, must have happened almost simultaneously since they are illuminating the same cloud. A single flare with a more complicated light curve could, in principle, describe this variability. Alternatively, the almost slowly variable emission could be associated with cosmic rays. X-ray observations in the next years will be crucial to discriminate between these scenarios. Indeed, as displayed in the third panel of Fig.~\ref{fig:spectral_variability}, the B.a region has reached a peak in the \Feka\ emission in the past $\sim 10$ years, and it is currently consistent with the measured flux at the beginning of the \xmm\ monitoring. If the \Feka\ emission is due entirely to the flaring activity of \sgras\ (or any other source), then in the next years the flux will drop, eventually below the value observed at the beginning of the \xmm\ observations.

\subsection{G0.11-0.11}

\begin{figure*}
    \centering
    \includegraphics[width=1.0\textwidth]{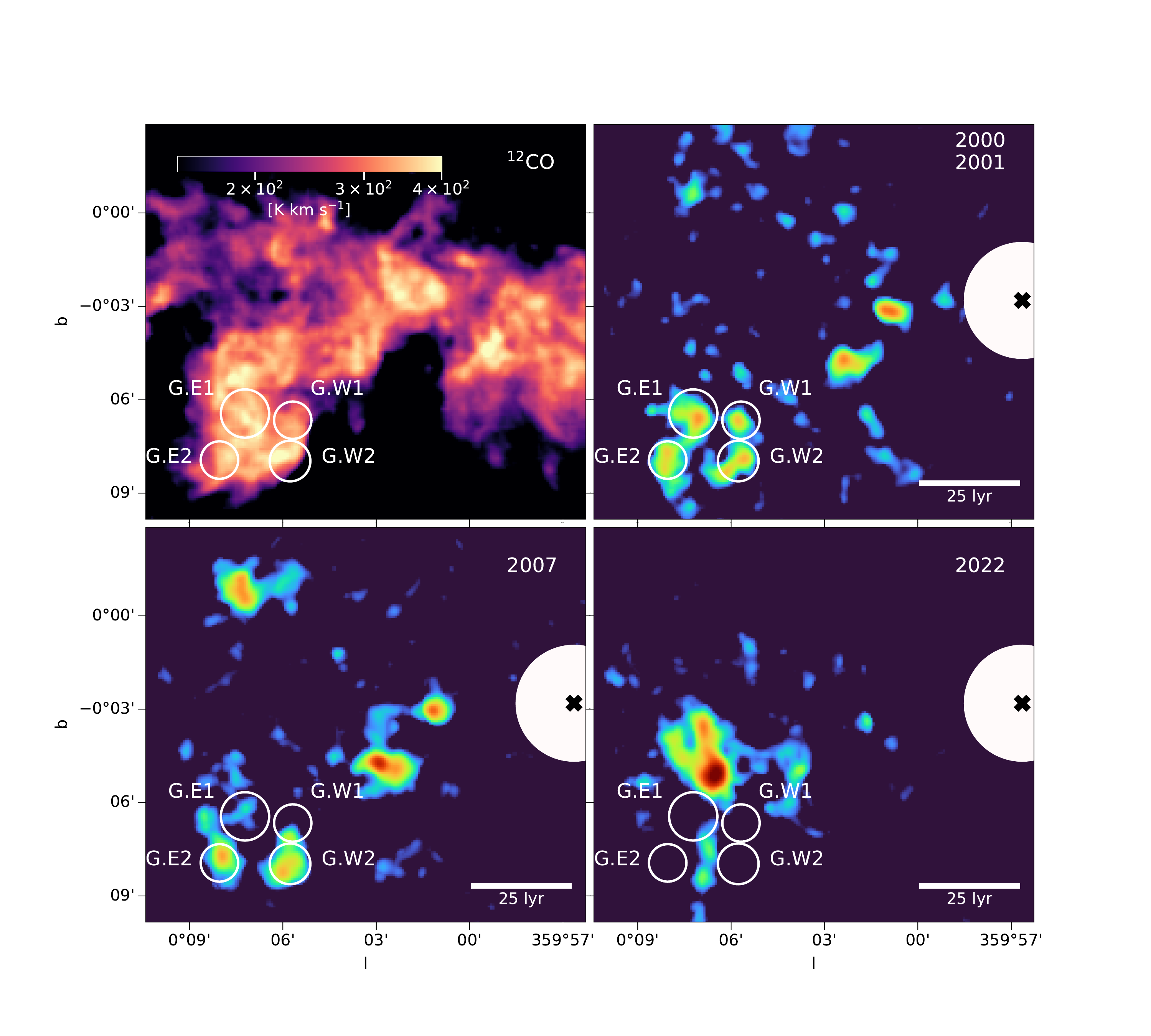}
    \caption{Fluorescence evolution in G0.11-0.11. The upper left image is the $^{12}$CO line intensity integrated in the range 25-45 km s$^{-1}$ \citep{Eden2020}. X-ray colorbar is consistent with Fig.~\ref{fig:variability_kalpha} and \ref{fig:variability_kalpha_sub}.}
    \label{fig:g011-011}
\end{figure*}

Details of the evolution of the cloud G0.11-0.11,  with a selection of three snapshots, are displayed in Fig.~\ref{fig:g011-011}. The upper left panel shows the $^{12}$CO line intensity in the velocity range 25-45 km s$^{-1}$. In the X-rays, it is possible to distinguish two separate ridges almost perpendicular to the Galactic plane. The western ridge is labeled as G.W in Fig.~\ref{fig:g011-011}, and, in the X-ray band, presents two substructures, i.e., G.W1 and G.W2.
The eastern ridge, similarly, appears to be structured as two distinct clumps (G.E1 and G.E2 in Fig.~\ref{fig:g011-011}). Fig.~\ref{fig:spectral_variability} (panel n.~4) shows the flux integrated on the 4 regions, as a function of time.

G0.11-0.11 was already illuminated in 2000. Therefore, also for this cloud, we can not know when the light front started to interact with it. 
Fig.~\ref{fig:variability_kalpha_sub} shows that the fluorescence signal monotonically decreases in the following years, the cloud then showed almost no residual variable \Feka\ emission after 2015. The variable component from the northern clumps (G.W1 and G.E1) was the first to vanish, followed by the southern part (G.W2 and G.E2).

Later on, an almost vertical feature between the previously discussed ridges started to appear as a bright \Feka\ emitter in  2019, staying bright until the end of our \xmm\ scan in 2024 (see Fig.~\ref{fig:g011-011} and Fig.~\ref{fig:variability_kalpha_sub}). This vertical feature is located at the same longitude as the B.b clump previously described in Sect.~\ref{sub:Bridge}, and it is aligned with the vertical tail that starts from B.b. The two clumps evolve on the same time scale, which can be an indication that they are a single coherent structure and the potential  connection channel between the two clouds. Indeed, high-resolution VLA observations of the G0.11-0.11 and the Bridge have shown that these two clouds are interacting \citep{Butterfield2022}.

\subsection{Arches Cluster and Bridge extension}

The molecular cloud associated with the position of the Arches Cluster is of particular interest since it has been proposed that it is interacting with the globular cluster. This cloud shows an almost constant emission until 2011 (Fig.~\ref{fig:variability_kalpha} and \ref{fig:variability_kalpha_sub}), as also suggested by the light curve in Fig.~\ref{fig:spectral_variability}. The initial non-variability of the \Feka\ was interpreted as a result of the bombardment of the molecular gas by cosmic rays accelerated by the shock induced by
the supersonic motion of the cluster through the interstellar medium. In particular, \citet{Tatischeff2012} studied in detail the interaction of this molecular cloud with low-energy cosmic rays. The authors concluded that the cloud can be illuminated by hadronic cosmic rays, accelerated in diffusive shock in the region of interaction of the Arches Cluster with the cloud. The proton power necessary to generate the observed \Feka\ flux in the early 2000s was $(0.2-1)\times 10^{39}$ erg s$^{-1}$. Whereas they estimated the kinetic power released in the collision between the star cluster and the molecular cloud to be $\simeq 2.3 \times 10^{40}$ erg s$^{-1}$. Such energy is sufficient to achieve the necessary power if the efficiency to accelerate particles is of the order of a few percent, which is expected in diffuse shock acceleration. They also ruled out the electron cosmic ray scenario. 

However, \citet{Clavel2014}, using \xmm\ data up to 2013, found an initial decrease of the \Feka\ emission, concluding that at least a significant fraction of the non-thermal emission is due to reprocessed hard X-ray radiation from the molecular cloud. Indeed, the timescale for diffusion or energy losses of ions requires at least decades \citep{Tatischeff2012,Clavel2014}; therefore, only interaction with  X-ray radiation can generate a  short time scale variability of the fluorescence. The drop of the \Feka\ intensity was confirmed by \citet{Krivonos2017,Chernyshow2018,Kuznetsova2019} using up to 2016 \xmm\ and \nustar\ data. \citet{Khabibullin2022}, using 2019 observations from SRG/eROSITA, found no clear detection of the 6.4 keV line, reporting an upper limit ($1 \sigma$) of $8\times10^{-10}$ cts s$^{-1}$ cm$^{-2}$ arcsec$^{-2}$ for a circular region with a radius of 50 arcsec. The same upper limit is found for a larger aperture of 100 arcsec radius. Our analysis confirms the strong decrease of the 6.4 keV emission (panel 2 in Fig.\ref{fig:spectral_variability}). The surface rate is roughly constant until 2013, at a level of $\simeq 3 \times 10^{-9}$ cts s$^{-1}$ cm $^{-2}$ arcsec$^{-2}$ in the 50 arcsec circular region shown in Fig.~\ref{fig:variability_kalpha} and Fig.~\ref{fig:variability_kalpha_sub}. In 2024, the line is barely detected, with a surface rate of $(8 \pm 2) \times 10^{-10}$ cts s$^{-1}$ cm $^{-2}$ arcsec$^{-2}$, therefore losing more than 70\% of the brightness in just a decade. The observed variability proves that the vast majority of the \Feka\ emission is due to reprocessed X-ray radiation from the molecular material. 

The globular cluster cannot be the illuminating source of this cloud. Indeed, the Arches Cluster is too faint in the X-ray band to produce the detected fluorescence in the early 2000s \citep{Tatischeff2012}, and even if some variability has been observed in the globular cluster, that amplitude is again too faint to generate the variable observed fluorescence \citep{Capelli2011}. Therefore, the cloud is most probably illuminated by an external source rather than the globular cluster. Given the proximity of the other bright clouds in the X-rays and a similar evolution in time, the cloud is likely illuminated by the same source, which illuminates the whole Sgr A complex.

The \Feka\ intensity has dropped by at least a factor $\sim 4$ during the 25-year period studied here (Fig.~\ref{fig:spectral_variability}). The residual, weak \Feka\ emission can still be associated with hadronic cosmic rays produced during the bow shock of the interaction of the cluster with the cloud. Future observations could provide tighter upper limits on the cosmic rays acceleration at the position of the cloud if the decreasing trend is confirmed in the next years.

We note that the Arches Cluster and G0.11-0.11 do evolve in a similar way. Moreover, since they are at about the same projected distance from \sgras, they should be located approximately at the same position along the LOS if the illuminating source is \sgras.

Some molecular material lies, in projection, between the \sgras\ and the Arches Cluster (Fig.~\ref{fig:median}). Coherently,  the X-ray emission also shows an excess in this region. This feature is visible in both Fig.~\ref{fig:median} and Fig.~\ref{fig:min}, where also the relative void between this feature and the Bridge can be observed. Such material has never shown a strongly enhanced \Feka\ emission (Fig.~\ref{fig:variability_kalpha_sub}), such as the clouds described before. Nevertheless, at least two clumps are visible in the minimum map (Fig.~\ref{fig:min}) between the position of \sgras\ and the Arches Cluster, with a correspondence in Fig.~\ref{fig:variability_kalpha}. The velocity of these clouds is different from that of the Arches Cluster, probably indicating that the two structures are not connected.

\subsection{50 km/s and 20 km/s molecular clouds}
\label{sec:variability:50_20}
In projection, the 20 km/s and 50 km/s clouds are closer to \sgras\ than the clouds we analyzed in the previous subsections. The 20 km/s cloud is located southwest of \sgras\, where no X-ray variability was observed in the last 25 years. Similarly, no variability was observed in a region consistent with the 50 km/s molecular cloud. In order to obtain an upper limit on the \Feka\ emission in these clouds, we collected a spectrum in a circular region of radius 1.3 arcmin, centered on the position of the 50 km/s ((l,b= 359.980, -0.071) deg). The radius is chosen to avoid contamination from the close MC1 cloud. For consistency, we use the same circular region centered at the position of the 20 km/s ((l,b= 359.887, -0.078) deg). We found an upper limit of $4 \times 10^{-6}$ photons s$^{-1}$ cm$^{-2}$ and of $1.5 \times 10^{-6}$ photons s$^{-1}$ cm$^{-2}$, respectively\footnote{The reported upper limit is the 90\% percentile of the posterior distribution of the \Feka\ flux.}. If these clouds are the closest to \sgras\, then the upper limits constrain the past flaring activity of the SMBH in the last century, as we will discuss in the next section.

In principle, these two giant molecular clouds can become bright in the future if an illuminating front from a different variable source reaches them. A potential candidate is the magnetar SGR J1745-2900, which underwent an outburst in 2013 \citep{Mori2013, Rea2013, Kennea2013}, and is known to orbit \sgras. We recently found the echo radiation of the outburst signal arising from the Circumnuclear Disk (CND), the clouds orbiting \sgras\ at a distance of 1.5 pc \citep{Stel2023}. If the 20 and 50 km/s molecular clouds are close and feeding the CND, and some dense knots lie between them, the wavefront of the magnetar will reach them in the future years and generate a fluorescence signal, detectable by \xmm\ if they are massive enough.

\section{The single flare scenario}
\label{sec:geometry}

The variability observed in Fig.~\ref{fig:variability_kalpha} and \ref{fig:variability_kalpha_sub} results from the interaction between the molecular material and the propagating wavefront of one or multiple past flares. In principle, any energetic outburst in the considered region can induce a fluorescence signal. 
 However, the natural candidate as the illuminating source is \sgras, and the polarization measurement of \ixpe\ has shown that the polarization angle of the reflected emission is consistent with this scenario \citep{Marin2023}. Therefore, we will restrict our analysis to \sgras, assuming that it is the illuminating source.
In this case, the spatial/temporal patterns of the \Feka\ glare are set by the relative position of the clouds with respect to \sgras. In the case of a single illuminating event, iso-delay curves are paraboloids with \sgras\ at the focus and the symmetry axis coinciding with the LOS. Starting from the flare event, the parabola scans the molecular cloud distribution. If $t$ is the elapsed time since the flare onset, and $R$ is the physical projected distance between a given point and the illuminating source, then the parabola's equation is \citep{Sunyaev1998}:
\begin{equation}
    z=c\frac{t^2-(R/c)^2}{2t},
    \label{eq:parabola}
\end{equation}
where $z$ is the distance from the source along the LOS.
Since the flare duration is presumably much shorter than the light crossing time of the Sgr A complex, the parabola scans only thin layers of the cloud distribution. For a short flare, in the optically thin limit, the X-ray surface rate (SR) is a direct probe of the molecular density in each layer illuminated by the wavefront \citep{Sunyaev1998}:
\begin{equation}
\begin{split}
    SR &= 7 \times 10^{-6} \left( \frac{n_{\mathrm{H_2}}}{10^4\, \text{cm}^{-3}} \right)  \left( \frac{\Delta t}{1 \text{yr}} \right) \left( \frac{L_8}{10^{39} \text{erg s}^{-1}} \right) \left( \frac{30 \, \text{pc}}{R} \right)^2 \times \\
    &\frac{\Omega}{(4.5 \, \text{arcsec})^2}  \frac{\eta^2}{1+\eta^2} \quad [\text{counts s}^{-1} \text{cm}^{-2} \text{ pixel}^{-1}],
\end{split}
\label{eq:SR}
\end{equation}
where $n_{\mathrm{H_2}}$ is the molecular hydrogen density, $\Delta t$ is the flare duration, $L_8$  is the source luminosity at 8 keV in an 8 keV wide energy band, $\Omega$  is the pixel size, and $\eta=R/(ct)$.
The velocity of the illuminating wavefront, along the LOS ($z$), can be derived by taking the time derivative of eq.~\eqref{eq:parabola}:
\begin{equation}
    v_{\rm los} \, (R,t)= \frac{c}{2} \left[ 1+ \left(\frac{R}{ct}\right)^2\right].
\end{equation}

In this section, we discuss the single flare scenario. We exploit the recent results of \ixpe, which pose the Bridge cloud 26 pc behind \sgras\ \citep{Marin2023}. This is equivalent to an age of the flare of 200 years, given the known projected distance of the Bridge from \sgras. Once for a given year, the position of the parabola is fixed in space by the polarization measurement, the position of the parabola for every other year is calculated using eq.~\eqref{eq:parabola}. Here, we aim to reconstruct the 3D distribution of the molecular material, and so the position of the individual clouds inside the Sgr A complex. Under the assumption of a 200-year-old flare, $v_{\rm los} \sim 0.2$ pc yr$^{-1}$. Note that the images we presented are temporally spaced by at least one year, which is similar to the thickness of the layer sampled by the parabola, assuming an upper limit on the flare duration of 1.5 years \citep{Clavel2013}. Moreover, the spatial resolution is slightly worse than the $\simeq 0.2$ pc sampling layer, but of the same order of magnitude\footnote{The pixel size is 0.18 pc at the Galactic center scale, with a $\simeq 10$ arcsec ($=0.4$ pc) PSF of the instrument.}.

\begin{figure*}
\centering
\includegraphics[width=1.0\linewidth]{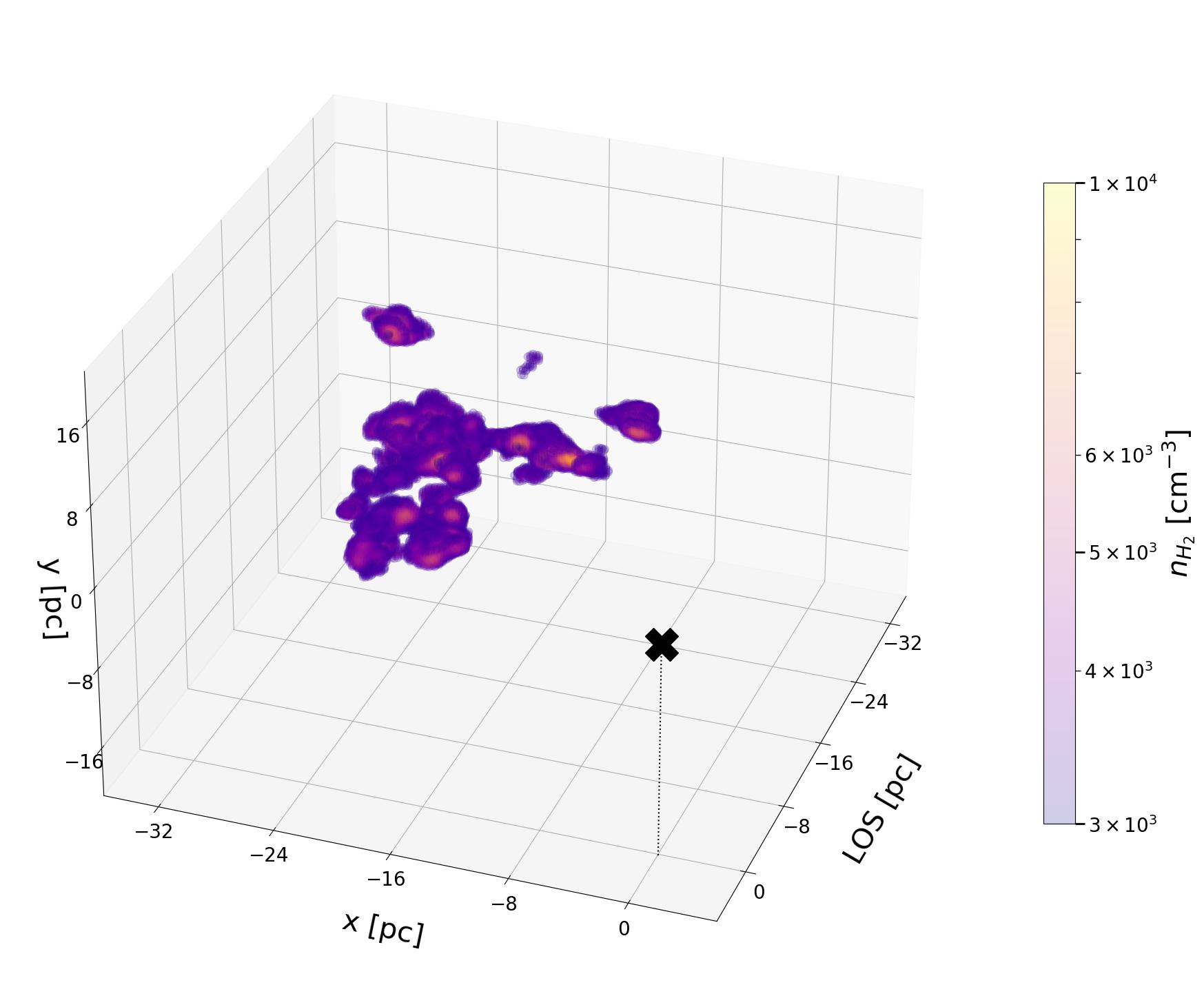}
\caption{Spatial distribution of molecular clouds in the Sgr A complex assuming they are illuminated by a 200-year-old flare from \sgras. \sgras\ is located in (0,0,0), a vertical dotted line is plotted to make its position at the axis origin more clear. y-axis points toward positive Galactic latitude; x-axis points toward negative Galactic longitude. Negative values along the line of sight (LOS) are behind \sgras.} 
\label{fig:clouds_distro}
\end{figure*}

We reconstructed the 3D distribution of H$_2$ density starting from the \Feka\ intensity maps (Fig.~\ref{fig:variability_kalpha}). For each observation, we considered the parabola of that given year and related the X-ray intensity of each point to the molecular density using eq.~\eqref{eq:SR}.
We assumed an X-ray luminosity of the flare of $L_{8}=10^{39}$ erg s$^{-1}$, a duration of 1.5 years, and that the flare happened 200 years ago. We also assumed that all the hydrogen is in molecular state and solar abundance for the clouds. Note that a different choice of the flare luminosity modifies the mean density of the derived Sgr A complex, while the relative densities are left unchanged, as it is clear from eq.~\eqref{eq:SR}.
The reconstruction of the distribution of the molecular clouds is presented in Fig.~\ref{fig:clouds_distro}. The black cross marks the position of \sgras. 
For graphical display purposes, we display in the map only regions with $n_{\rm H_2}> 3 \times 10^3$ cm$^{-3}$.

\begin{table}[ht]
    \caption{Angular separation, projected distance at 8.2 kpc, and estimated LOS distance from \sgras, assuming a single outburst event happened 200 years ago.}
    \centering
    \begin{tabular}{c|c|c|c}
    \hline
    Cloud & \makecell{Angular \\ separation}  &  \makecell{Projected \\ distance} & \makecell{Estimated \\ LOS distance}\\
          & [arcmin]            & [pc] & [pc]\\
    \hline
    MC1  & 4.4  & 10.5 & 27.4\\
    MC2  & 6.0 & 14.3 & 26.0\\
    Arches Cl. & 11.4 & 27.2 & 21.7\\
    Bridge a & 7.6 & 18.1 & 25.7\\
    Bridge b & 10.0 & 23.9 & 25.9\\
    G 0.11-0.11 & 11.6 & 27.7 & $\le$ 20\\
    \hline
    \end{tabular}
    \label{tab:separation}
\end{table}

\begin{figure*}
\centering
\includegraphics[width=1.0\textwidth]{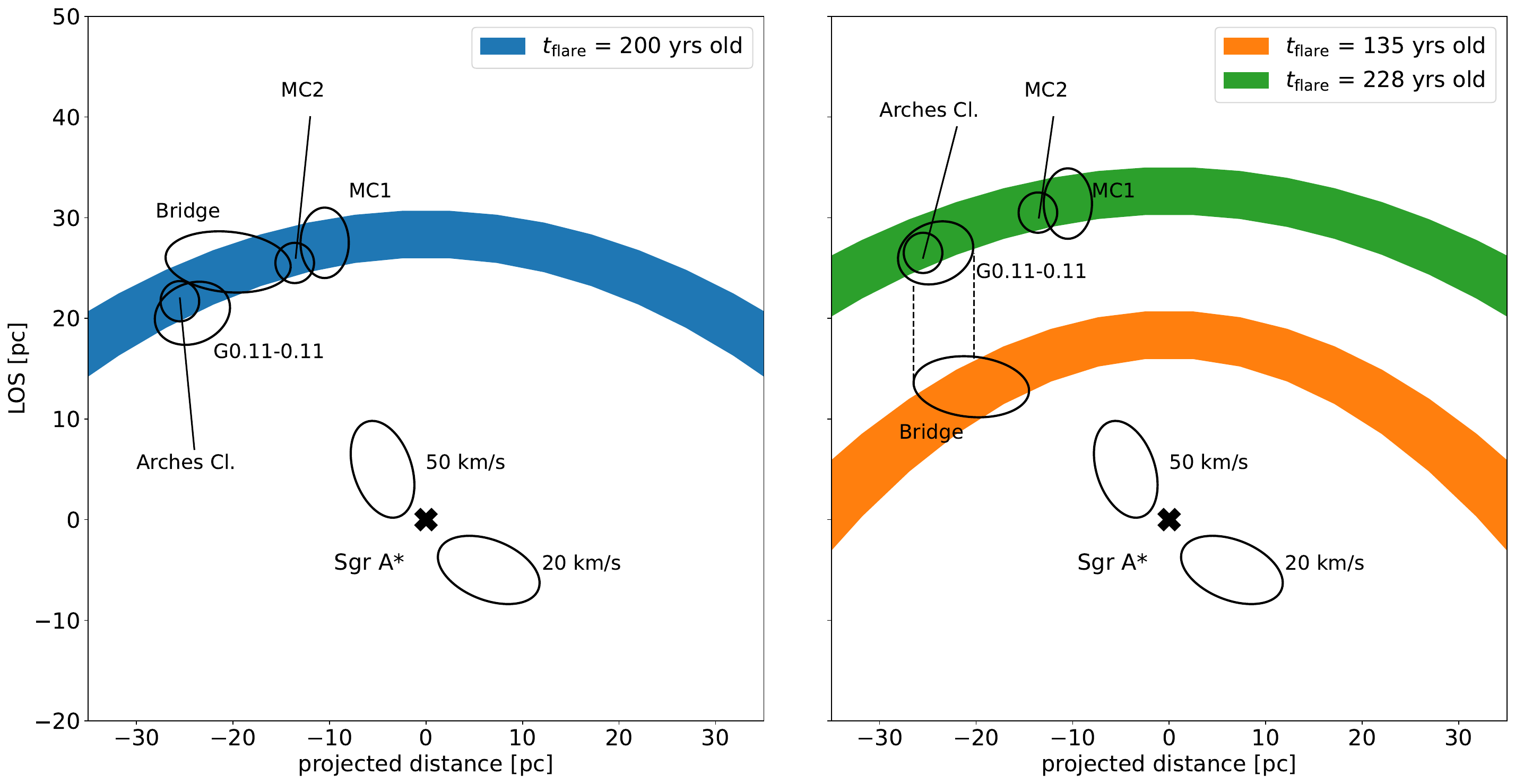}
\caption{Top view of the molecular clouds in the Sgr A complex. {\it Left panel:} single flare scenario. The Sgr A complex appears as a compact structure located $\simeq 25$ pc behind \sgras. The width of the parabola refers to the space sampled by the wavefront in the last 30 years, rather than the duration of the flare. {\it Right Panel:} multiple flares scenario described in \citet{Clavel2013} and \citet{Chuard2018}. The Sgr A complex is illuminated by a short flare (happened $\simeq 135$ years ago) and a longer flare ($\simeq 228$ years ago). Two dashed lines highlight the connection between the Bridge and G0.11-0.11.} 
\label{fig:top_view}
\end{figure*}

\begin{figure*}
\centering
\includegraphics[width=1.0\textwidth]{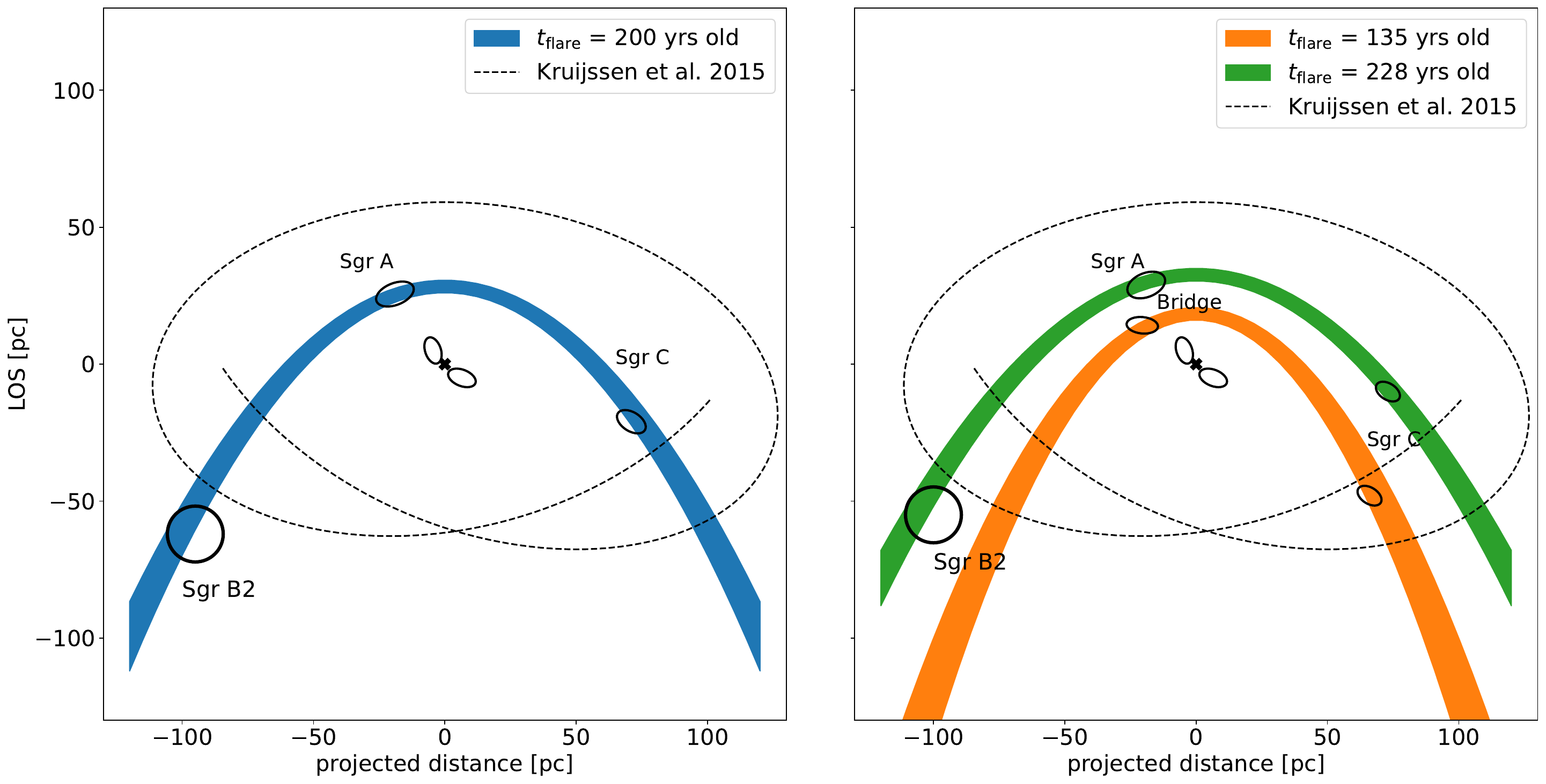}
\caption{Similar as Fig.~\ref{fig:top_view}, but the whole CMZ region is shown. The dashed line is the dynamical model proposed by \citet{Kruijssen2015}. Most of the molecular clouds in the CMZ should roughly lie on the ellipsoid. Regardless of the scenario, the Sgr A complex appears to be located at smaller radii than the nuclear ring.} 
\label{fig:top_view1}
\end{figure*}

\subsection{3D distribution}
Fig.~\ref{fig:top_view} (left panel) displays a top-view of the reconstructed geometry of the Sgr A complex under the assumption of a single flare happened 200 years ago. The right panel of the same figure shows the multiple flares scenario, which will be discussed in Sect.~\ref{sec:multiflare}. The area sampled by the flare in the past 30 years of high-resolution observations in the X-ray band is highlighted in blue in the same figure. The Sgr A complex appears as a compact group of molecular clouds, located $\simeq 25$\,pc in the background of \sgras. The parabola has moved $\simeq 5$\ pc along the LOS in the last 25 years. Given the distribution of the cloud, a $\simeq 10-15$ pc region has been illuminated in this direction. 
In Table~\ref{tab:separation}, we report the separation and the estimated distance along the LOS from \sgras\ of each cloud. Note that since the maximum \Feka\ intensity of G0.11-0.11 is already present in the first observation, for this cloud, we can only give an upper limit to the distance from \sgras. 
The position of the Sgr A complex along the LOS depends on the age of the flare. For example, if the flare happened 100 years ago, the Sgr A complex would be closer to \sgras\, with the clouds distributed within 0 and 15 pc along the LOS from the SMBH. Whereas, if the age of the flare is set to 400 yrs, the molecular complex would be about 60 pc behind \sgras.

The exact position of several molecular clouds in the CMZ is controversial \citep{Henshaw2023,Walker2024,Lipman2024}. Most molecular clouds should lie on the molecular nuclear ring encompassing the CMZ. The ring is formed by accreting molecular material toward the center of the Galaxy, in the gravitational potential dominated by the bar. Different kinematical models result in slightly different shapes of the ring. Nevertheless, in all kinematic models, the clouds lie in a range between 50 and 200 pc from \sgras. For example, in \citet{Kruijssen2015}, different clouds lie on different open streamers. In Fig.~\ref{fig:top_view1}, we plotted the streamers of this particular dynamical model (dashed lines), along with the position of the Sgr A complex and the space sampled by the parabola in the single flare scenario. In the dynamical model, most of the clouds of the CMZ should lie on the dashed lines. In the same figure, we reported the position of two other molecular complexes: Sgr B2 and C, in the case the same flare illuminates them. Interestingly, the position of Sgr B2 is consistent with the one derived by \citet{Kruijssen2015}, given the size of the cloud and the uncertainty on the exact age of the flare. If the Sgr A complex belongs to one of these streamers, it should be approximately no less than 50 pc in front or behind \sgras. {This is not the case we infer from the \ixpe\ measurement of the flare age. For example, to be located roughly 50 pc behind \sgras, the age of the flare should be about 400 years.}
In this paper, we argue that if a single 200-year-old flare illuminates the whole Sgr A complex, then it is not part of the ring of molecular clouds that forms the CMZ, but rather, inside the ring. The CMZ would then be more complicated than the state-of-the-art models that describe the dynamic of molecular clouds in the CMZ. The molecular complex could be part of the material falling from the ring toward the inner parsecs of the Galactic center. 

\begin{figure*}
    \centering
    \includegraphics[width=1.0\textwidth]{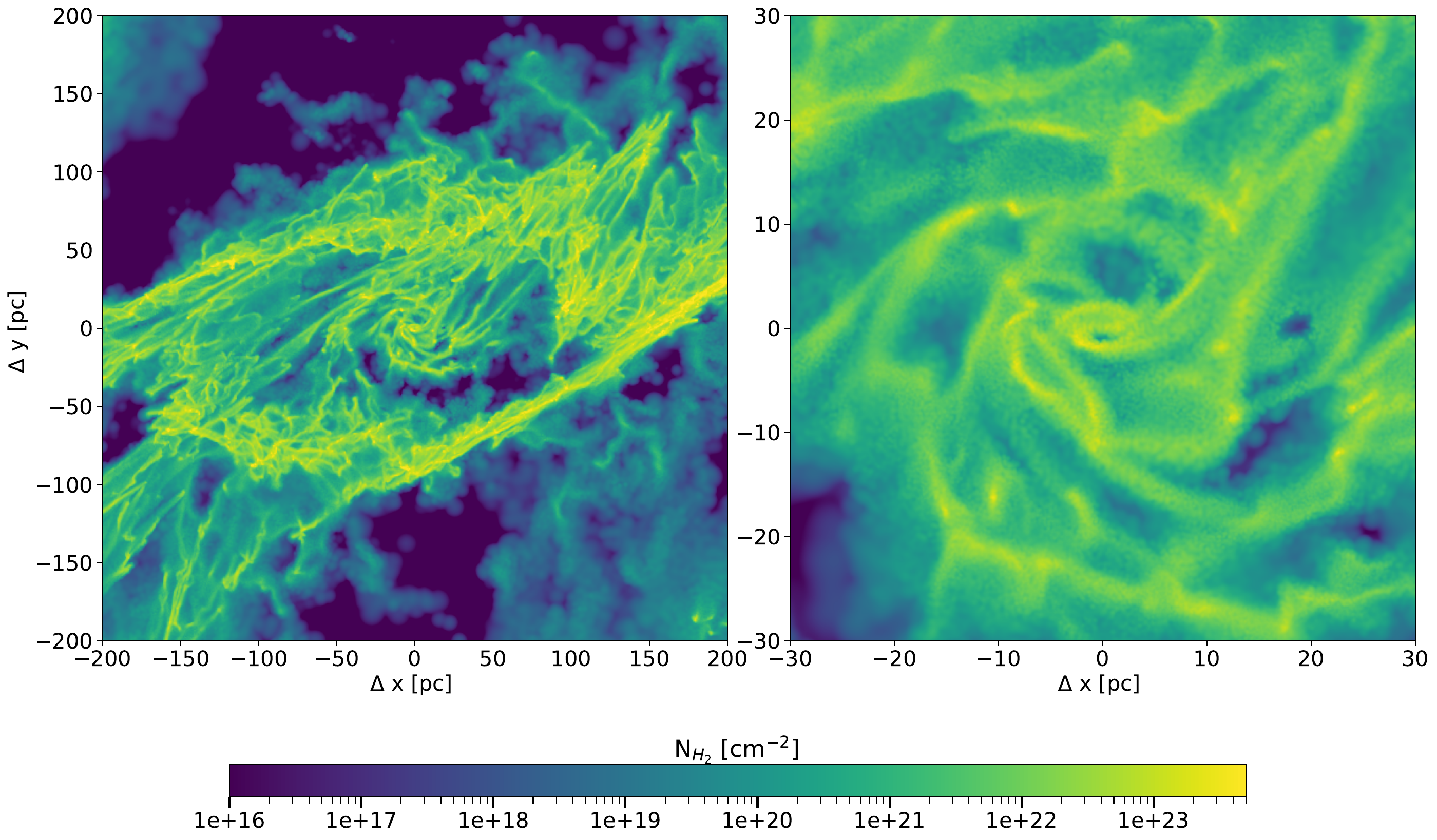}
    \caption{Top view projection of the molecular hydrogen column density in the simulation box studied in \citet{Sormani2020}. {\it Left panel:} inner 200 pc of the simulated box, the ring is the analog of the CMZ of the Milky Way. {\it Right panel:} the inner 30 pc of the simulation. Some streamers are visible, feeding the Galactic center with molecular gas from the molecular ring.}
    \label{fig:cmz_sim}
\end{figure*}

Similar streamers of material are observed in 
the core of outer galaxies \citep{Sun2024} as well as in 
hydrodynamical simulations. Fig.~\ref{fig:cmz_sim} displays an example of a simulation box studied in \citet{Tress2020,Sormani2020}. The goal of the simulations presented in those works is to study gas flows and star formation in the CMZ, while also taking into account the strongly non-axisymmetric large-scale flow driven by the Galactic bar, in which the CMZ is embedded, out to R $\sim5$ kpc. The simulations include a realistic Milky Way external barred potential, a physically motivated sub-grid prescription for star formation and supernova feedback through the use of sink particles, a three-phase ISM (cold, warm, hot), and a non-equilibrium chemical network that can keep track on the fly of several chemical species (H, H+, H2, CO, C+ - basically hydrogen and carbon chemistry). The simulations reach sub-parsec resolution in the dense regions (such as CMZ molecular clouds and CND) and self-consistently follow the formation of a CND-like structure from the large-scale flow. The left panel shows a typical nuclear ring analog to the CMZ of the Milky Way. The right panel is a zoom-in of the inner 30 parsecs. The zoom-in shows some molecular material drifting toward the inner parsecs of the Milky-Way, in stretched and elongated arched features. The Sgr A complex could be similar to one of the stretched streamers present in the figure. The reconstructed geometry of the Sgr A complex presented in Fig.~\ref{fig:top_view} may also remember the arched shape of the infalling molecular gas in the simulation. However, this similarity is related to the shape of the parabola, which illuminates the molecular cloud distribution.

\section{Probability density function of the molecular density in the single flare scenario}
\label{sec:pdf}

\begin{figure}
\centering
\includegraphics[width=1.0\linewidth]{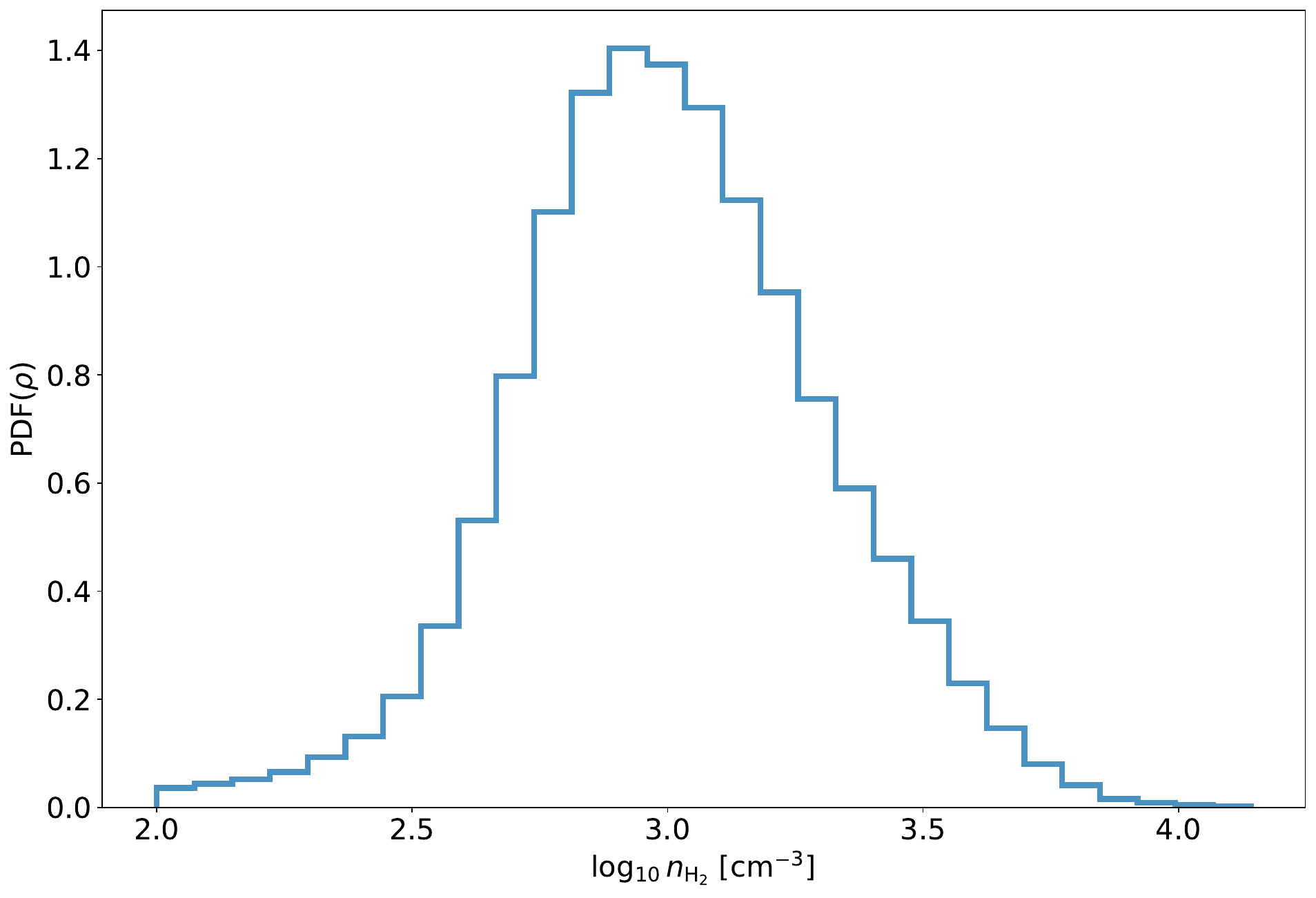}
\caption{\pdfrho\ of the illuminated part of the Sgr A complex in the last 25 years.} 
\label{fig:pdf}
\end{figure}

Fig.~\ref{fig:pdf} shows the derived PDF of the H$_2$ number density of the Sgr A complex (\pdfrho, blue histogram). 
The molecular gas density PDF is a fundamental tool to understand the evolution of the gas in the Galaxy since it is directly related to the star formation rate \citep{Hennebelle+2012, 
Hennebelle2013, Padoan+2014}. 
The shape of \pdfrho\ is mainly determined by the supersonic turbulent flow of the molecular gas. If $s \equiv  \ln(\rho/\rho_0)$, where $\rho_0$ is the mean density, then $s$ is expected to be normal distributed \citep{Vazquez-Semadeni+1994,Padoan+1997} with the width of the distribution which follows:
\begin{equation}
    \sigma_s=\ln{(1+b^2\mathcal{M}^2)},
\end{equation}
where $\mathcal{M}$ is the Mach number, and $b$ is a dimensionless parameter that depends on the mode of the turbulence. It ranges from $b=1/3$ for pure solenoidal turbulence to $b=1$ for a pure compressive mode \citep{Federrath+2008}.
A purely turbulent distribution develops a power law tail at high density in the case of self-gravity, and, therefore, star formation \citep{Federrath+2013, Ballesteros-Paredes2011, Appel2023, Veltchev2024}. Magnetic fields can further modify the shape of the PDF as a second-order effect \citep{Federrath+2012}.
The \pdfrho\ has been studied in molecular radio surveys for several cloud complexes in the Milky Way. Log-normal shapes, both with and without power-law tails, have been observed \citep{Schneider+2015, Alves+2017, Spilner+2021, Ma2022, Murase2023,
Kinman+2024}. As already said, X-ray observations allow us to derive the \pdfrho\ only for the molecular clouds that are directly illuminated and only in the case of a short flare. On the other hand, the reflection signal directly provides the local density of the molecular gas, whereas molecular radio surveys are sensitive to the integrated column density. The \pdfrho\ has been measured with 2015 \chandra\ data \citep{Churazov2017c}. The authors found that this dataset roughly follows a log-normal distribution with a width of $\sigma_s \simeq 0.7$. Here we study how the density is distributed by looking at the entire \xmm\ monitoring of the region to verify if adding 15 more "snapshots" provides a consistent result with the one presented in \citet{Churazov2017c}.

\begin{figure}
\centering
\includegraphics[width=1.0\linewidth]{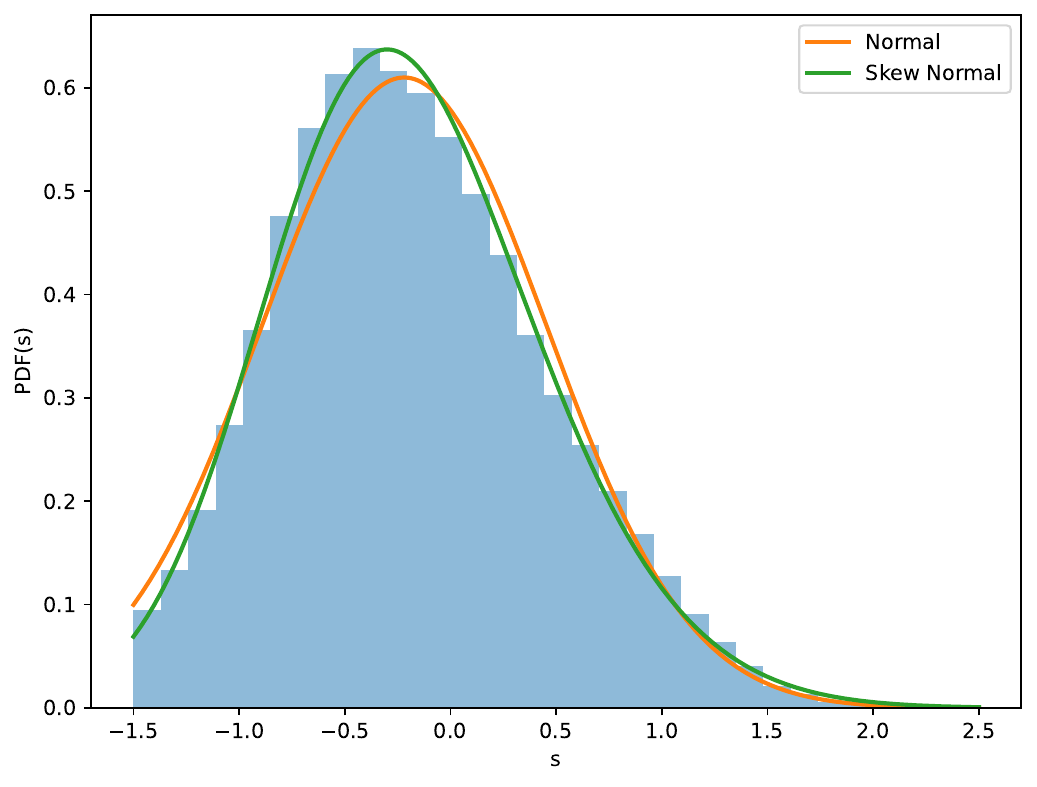}
\caption{Fit to the distribution of the $s$ variable (the density normalized by its mean value). The fit does not depend on the binning choice. Both fitting functions are truncated at $s*=-1.5$.} 
\label{fig:pdf_fit}
\end{figure}

\begin{figure}
\centering
\includegraphics[width=1.0\linewidth]{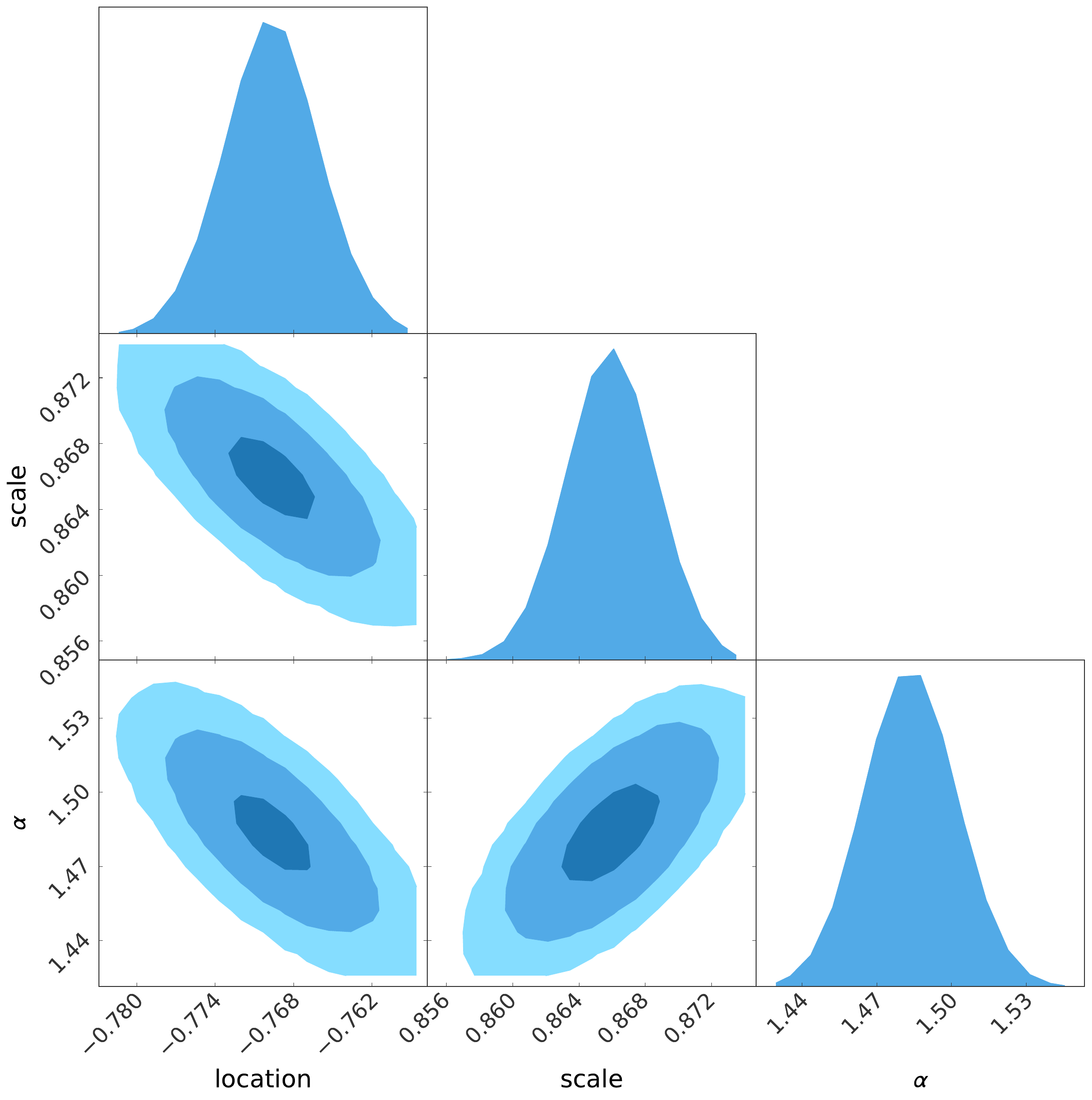}
\caption{Posterior distributions for the parameters of the truncated skew normal for the \pdfs\ shown in Fig.~\ref{fig:pdf_fit} } 
\label{fig:posterior}
\end{figure}

We first perform a fit with a truncated normal function. The fit is performed via Markov-Chain-Monte-Carlo (MCMC) on the unbinned distribution of data shown in Fig.~\ref{fig:clouds_distro}, using the \text{emcee} python package \citep{Foreman-Mackey2013}. Since the distribution is noisy at the low-density end, we decided to truncate the distribution downwards at an arbitrary value $s*$.
Indeed, background and poissonian noise affect the low-density tail of the distribution. The second arises from the intrinsic Poissonian distribution of the incoming photons collected by the telescope. This uncertainty is higher where the surface brightness (and, therefore, the reconstructed density) is lower. The orange line in the histogram (Fig.~\ref{fig:pdf_fit}) displays the best-fit distribution, assuming $s*=-1.5$ as the lower limit of the PDF. The histogram presents a clear asymmetry that the fitting function struggles to reproduce. Moreover, we tested that different choices of the truncation value return different and inconsistent best-fit values for the parameters. We then performed a second fit using a truncated skew normal function (green line in Fig.~\ref{fig:pdf_fit}). The skew normal distribution generalizes the normal distribution by allowing a non-null skewness. The skewness is controlled by the shape parameter ($\alpha$). For a right-tailed distribution $\alpha > 0$, while  $\alpha = 0$ represents the normal distribution case. The skew normal distribution is defined as:
\begin{equation}
    \mathrm{PDF}(x,\alpha) = 2 \phi (x) \Phi (\alpha x),
\end{equation}
where $\phi$ is the normal distribution and $\Phi$ is its cumulative distribution function. To add the usual location and a scale parameter, the PDF is modified accordingly to:
\begin{equation}
    \mathrm{PDF}(x,\alpha) \rightarrow \frac{1}{\mathrm{scale}}\mathrm{PDF} \left( \frac{x-\mathrm{location}}{\mathrm{scale}},\alpha \right)
\end{equation} 
In this case, the posteriors of the parameters do not depend strongly on the particular choice of the truncation value, making the analysis sound. Fig.~\ref{fig:posterior} shows the corner plot of the fit. The shape parameter is  strictly constrained $\alpha > 0$, suggesting that the \pdfrho\ differs from a pure log-normal behavior at the high-density end\footnote{ The same behavior is confirmed by using EPIC-mos data, suggesting that it is not a side effect of the construction of the maps. The $\alpha$ parameter is strictly constrained to be positive and consistent within 3 $\sigma$.}. In all cases, the distribution width is $\sigma_s \simeq 0.7$, consistent with the value found by \citet{Churazov2017c} using 2015 Chandra data.

It must be noted that the \pdfrho\ derived from the X-ray variability presents some limitations and is affected by many systematics \citep{Khabibullin+2020}. Background noise is present at the low-density end of the distribution, even if we reduce its contribution by truncating the distribution and looking at a narrow energy range of energy in which the reflection component is dominant. Furthermore, the exposure of each image defines the Poissonian noise of the reflection component. Since we observe variability among maps that are one year apart in time, the Poissonian noise can not be reduced without losing information on the temporal variability of the signal. Moreover, at high density, the PDF is affected by the instrumental resolution (PSF), the smoothing process, and the thickness of the parabola that is illuminating the clouds. Indeed, both effects can be larger than the typical size of the smallest and more dense clumps. Further limitations are due to the fact that the \pdfrho\ refers only to the region of the Sgr A complex that has been illuminated in the last 25 years. Since the clouds were already illuminated in the year 2000 and are still on in the latest observations, other parts of the complex can not contribute to the density distribution, especially at high densities where clouds are intrinsically rarer (sample variance effect). Moreover, the density information in the layers corresponding to the years in which no observations were taken, such as 2017 or 2018, is irremediably lost. Finally, one should keep in mind that our analysis was performed in the optically thin limit, whereas in the more dense region of each cloud, the reflected signal does not scale linearly with the investing flux because of the opacity of the cloud.

\section{Multiple flares scenario}
\label{sec:multiflare}
\citet{Clavel2013} proposed that two different flares from \sgras\ are currently illuminating the Sgr A complex. Indeed, they observed a single filament in the Bridge, in which the X-ray reflection evolves on a time scale of $\sim$ 2 years, contrasting with the longer evolution time scales ($\sim 10$ yrs) observed in G0.11-0.11, MC1, and MC2. However, it is worth noting that \citet{Muno2007} reported a shorter time variability (2-3 years) in MC1 and MC2. Moreover, \citet{Clavel2013} identified a significant difference in the ratio of the X-ray intensity to the peak temperature (in Kelvin) measured in the molecular line N$_2$H$^{+}$, which is proportional to the column density, between the Bridge and the other clouds. Specifically, while the clouds exhibit similar \Feka\ fluxes, the Bridge shows a temperature peak signal that is 2-3 times higher than the other clouds. If a single short flare illuminates the whole complex, then the X-ray intensity should roughly scale with the temperature peak signal (which is a proxy for the molecular column density). Therefore, to reproduce similar X-ray brightness, the denser Bridge must reflect a different flare than the other clouds. 
This evidence leads the authors to propose a multiple flares scenario, in which a short flare illuminates the Bridge, whereas the remaining clouds are reflecting a much longer ($\gtrsim 10$ yrs) energetic event. In particular, the Bridge would be illuminated by a short fraction along the LOS. In contrast, the remaining less dense clouds would be illuminated for a larger portion along the LOS.
The multiple flares scenario was further developed by \citet{Chuard2018} studying the spectra of different clumps in the Sgr C complex. They found that the time variability observed was better described by two distinct flares separated by about 100 years. In particular, they suggested that the ages of the two flares are about $162^{+27}_{-17}$ (the short one) and $267^{+20}_{-25}$ years old (the long flare), assuming solar metallicities, and 2024 as the reference year. For higher abundances (Z=1.3) the flare ages are $135^{+14}_{-11}$ and $228^{+24}_{-16}$ years. The authors claimed that the two flares that illuminate the Sgr C complex may be those indicated in the scenario proposed by \citet{Clavel2013} for the Sgr A complex.

The long monitoring described in this work adds crucial information to the proposed scenario. For example, the western part of the Bridge (B.b) has significantly become brighter by at least a factor of $\sim 3$, with respect to the first decade of observation. It reaches a surface brightness significantly higher than the remaining clouds in the Sgr A complex (Fig.~\ref{fig:variability_kalpha}). This measure overcome the fact that in the early 2000s the denser Bridge showed a similar \Feka\ flux of less dense clouds. This fact alone does not rule out the possibility that the Sgr A complex is responding to two different flares, but it solves the main weakness of the single-flare scenario. Indeed, in a single and short flare scenario, one expects the most dense knots to be the brightest in the X-rays. The presented observations reproduce this behavior.
Moreover, the 25 years of \xmm\ observations analyzed in this work show that also the Bridge, at least at the scale considered in this work, presents a long time evolution (e.g., B.b), similar to the one observed in other clouds, as can be appreciated in the light curves presented in Fig.~\ref{fig:spectral_variability}. This fact could indicate that the evolution of the different time scales is indeed due to the internal structure of the molecular gas in the complex (presence of clumps and envelopes).

The right panel of Fig.~\ref{fig:top_view} depicts a top view of the cloud disposition and the regions illuminated by the parabolas in the multiple flares scenario developed by \citet{Clavel2013}. Here, we assumed the age of the flares measured by \citet{Chuard2018} in the Sgr C complex, in the case of Z=1.3 solar metallicity. 
In this scenario, the Bridge structure would be much closer to \sgras\ (roughly 10-15 pc behind it), whereas the remaining cloud would be $ 25-30$ pc behind \sgras.
The Bridge and G0.11-0.11 are separated by $\sim 10$ pc along the LOS. The fact that they are interacting \citep{Butterfield2022}, in the figure highlighted by dashed lines, cannot completely exclude the multiple flares scenario. Indeed, the reported separation between the two clouds is comparable to the typical size of these clouds. However, if this is the case, the vertical structure we observed in G0.11-0.11 cannot be the connection between this cloud and the Bridge. In fact, the short-time variability observed in this region cannot be associated with a 10 pc long filament.

Even if, in the multiple flares scenario, the clouds of the Sgr A complex span a larger distance along the line of sight, they are still at smaller radii than the nuclear ring. In the right panel of Fig.~\ref{fig:top_view1}, we displayed the multiple flare scenario for the whole CMZ. As concluded in Sect.~\ref{sec:geometry}, the Sgr A complex may be drifting from the nuclear ring to the inner parsecs of the Galaxy. In this case, however, the cloud distribution is more fragmented than the more compact and arches-like one obtained in the single flare scenario (see Fig.~\ref{fig:top_view}). Sgr B2 is roughly consistent with the position derived in \citet{Kruijssen2015}. Additionally, in the geometry described by \citet{Chuard2018}, Sgr C is also not too distant from the ellipsoid proposed by \citet{Kruijssen2015}; however, it is formed by two different clumps separated by about 40 pc along the LOS.
If the scenario depicted in the right panel of Fig.~\ref{fig:top_view1} is correct, then Sgr B2, which is already been swept by the older flare, will be illuminated by the second, more recent, flare. Given the size of Sgr B2 ($\sim 40$ pc), this second wavefront should reach the massive molecular cloud about 30 years from now. 

Note that the \pdfrho\ derived in Sect.~\ref{sec:pdf} is no longer valid in the multiple flares scenario. Indeed, only in the case of a short flare, the region illuminated is a narrow parabola, and the X-ray brightness is a probe for the molecular density. In the case of a longer flare, each cloud is illuminated in a larger region along the LOS. The X-ray emission would then scale with the column density of the cloud if the duration of the flare is long enough to illuminate the whole cloud. Therefore the results shown in Sect.~\ref{sec:pdf} are correct only in the single flare scenario.

The final proof of the interaction of the molecular complex with two different wavefronts would be any repeated or somewhat cyclic behavior of the  \Feka\ light curve in any of the clouds analyzed in the present work. In this scenario, the clouds would become bright by interacting with the first flare, then the fluorescence would drop as the first wavefront passes the cloud distribution, and they would re-bright again as the wavefront of the second flare reaches them. This kind of signal has not been observed so far in the 25-year \xmm\ monitor of the region. That period can be extended to 30 years by considering the observations of ASCA and Beppo-Sax in the '90s, in which G0.11-0.11 was the brightest cloud of the complex. Therefore, if the Sgr A complex had been illuminated by two different energetic events, this would have happened at least 30 years apart.

\begin{figure}
  \centering
  \includegraphics[width=\linewidth]{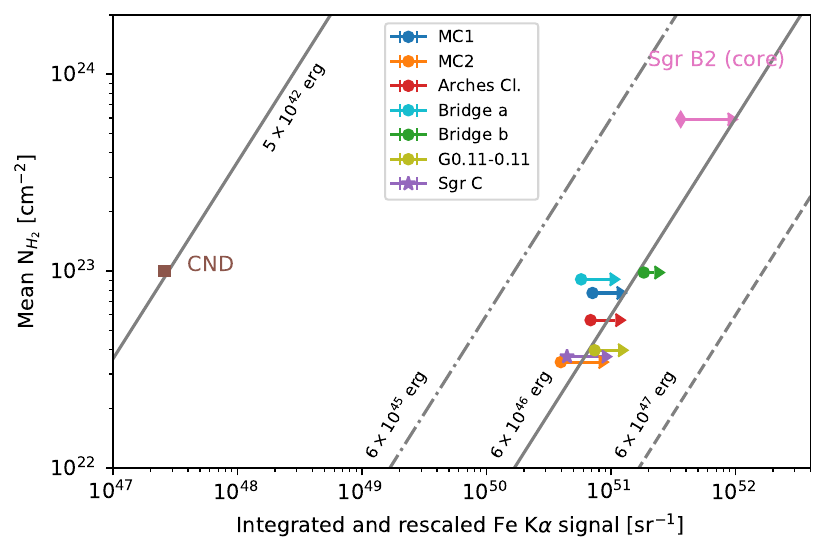}
  \caption{Comparison between the molecular hydrogen column density ($N_{\mathrm H_2}$) and the observed \Feka\ signal of the illuminated clouds in the CMZ. The \Feka\ signal represents the time-integrated observed surface rate (i.e. the area under the light curves shown in Fig.~\ref{fig:spectral_variability}), rescaled by multiplying it by the square of the physical distance ($R^2$) between the cloud and \sgras. The lines show the theoretical predictions for different values of total emitted energy in the 2–10 keV band. The $N_{\mathrm H_2}$ values are derived from the map published by \citet{Battersby2024}. Data for Sgr B2 and Sgr C are taken from \citet{Terrier2010,Terrier2018,Chuard2018}. Data for the CND, from \citet{Stel2023}, reflect its response to the significantly weaker flare from the magnetar SGR J1745-2900.}
\label{fig:nh2_fluence}
\end{figure}

Although our analysis revealed that the densest clouds in the Sgr A complex are the brightest ones, it is worth extending the comparison between column density and the measured \Feka\ signal to the clouds illuminated across the entire CMZ. This comparison is illustrated in Fig.~\ref{fig:nh2_fluence}, which also includes the expected relations for different values of the total energy emitted by the putative \sgras\ flare\footnote{See eq.~(4) in \citet{Sunyaev1998} and eq.~(A.3) in \citet{Stel2023}}. The hydrogen column density is from far-IR dust maps \citep{Battersby2024}, with additional X-ray data for Sgr B2 \citep{Inui2009, Terrier2010,Terrier2018}, Sgr C \citep{Chuard2018}, and the CND \citep{Stel2023}. The CND, illuminated by the flare of the magnetar SGR J1745-2900 which occurred in 2013, but not by past \sgras flares, exhibits a \Feka\ signal orders of magnitude weaker than those of the Sgr A complex. This is due to the magnetar's significantly lower total energy output compared to the \sgras\ past flare(s). 

In the case of a short flare that illuminates only thin layers of the molecular clouds, the contributions from multiple years of observations must be summed to obtain a signal proportional to the column density. Accordingly, the signal reported on the X-axis of Fig.~\ref{fig:nh2_fluence} represents the integral under the light curves shown in Fig.~\ref{fig:spectral_variability}. However, these light curves are incomplete, as some \Feka\ line emission was already present prior to the onset of the monitoring period, and additional emission is anticipated in the coming years. Consequently, the reported values should be regarded as lower limits. The actual values are estimated to be up to 2–4 times higher, given that the illuminated fraction of some clouds corresponds to only 25–50\% of the typical cloud size (see Sect.~\ref{sec:geometry} and Fig.~\ref{fig:top_view} and \ref{fig:top_view1}). Additionally, the signal is rescaled by a factor of $R^2$, where $R$ is the physical distance between the cloud and \sgras\ as derived in Sect.~\ref{sec:geometry}. Notably, the adoption of $R$ values derived from the multiple flares scenario (right panel of Fig.~\ref{fig:top_view}) introduces a negligible impact on the overall results.

As shown in Fig.~\ref{fig:nh2_fluence}, the data points for the illuminated clouds are in agreement with the prediction for a flare that released a total X-ray energy of $6 \times 10^{46}$ erg (corresponding to an X-ray luminosity of $10^{39}$ erg s$^{-1}$ over a duration of two years). The dashed and dash-dotted lines represent the theoretical predictions for a flare with 10 times higher and 10 times lower energy output, respectively. The clustering of data around the same theoretical prediction, with a scatter of a factor of 2–3, implies that multiple flares, if present, must have emitted comparable amount of energies within a similar factor.

\section{Constraints from the 50 and 20 km/s molecular clouds}
\label{sec:20-50}
There is evidence of streamers arising from the 20 km/s and 50 km/s clouds feeding the inner parsecs of the Galactic center, connecting these two giant molecular clouds with the clouds orbiting \sgras\ at a distance of 2 pc \citep{Ho1991, McGary2001, Hsieh2017}. The clouds orbiting \sgras\ on this scale form the CND \citep{Christopher2005, Requena-Torres2012, Lau2013, Hsieh2021}. The physical connection between the CND and the two giant clouds makes them the closest to the Galactic center. Notably, the 20 km/s is most probably situated slightly in front of \sgras, while the 50 km/s is slightly behind it \citep{Coil2000, Herrnstein2005, Ferriere2012}. Moreover, the 50 km/s is interacting with the supernova remnant Sgr A East, which is a few pc behind \sgras\, \citep{Ehlerova2022}. These observations constrain the position of the two massive clouds within 10-20 pc of \sgras, even if some dynamical models place them differently \citep{Henshaw2023}. If the described geometry is correct, then the wavefront of any \sgras\ flare must have already swept through this region. For example, if a single flare happened 200 years ago, as suggested by the \ixpe\ measurement, then these clouds were illuminated between the epoch of the flare and the first high-resolution observation by ASCA in 1993. 

The upper limit on the \Feka\ intensity, measured in Sect.~\ref{sec:variability:50_20}, places an upper limit on the past \sgras\ luminosity. In particular, the 50 km/s cloud, being behind the illuminating source, should reflect the light for a longer period of time. In fact, the back farthest part of the cloud is the last to be illuminated since the light has to travel to this part and then reach the observer. If the cloud is located according to the described geometry, then this last part to be illuminated would be about 10 pc behind \sgras\ along the LOS and 5 pc in the projected direction across the sky from it. Using eq.~\eqref{eq:parabola}, this position translates into a 70-year time delay. Note that the first high-resolution observations of the Galactic center were performed about 30 years ago with ASCA, where no evidence of \Feka\ radiation from 50 km/s cloud was detected. Therefore, the outburst of \sgras\ should have happened no less than $\sim 100$ years ago. Moreover, the upper limit on the intensity can be translated  to an upper limit of the X-ray luminosity \citep{Sunyaev1998}. Assuming a mean hydrogen column density of $N_{\mathrm H_2}=10^{23}$ cm$^{-2}$ \citep{Ponti2010}, the X-ray luminosity of \sgras\ is constrained $L_X < 10^{36}$ erg s$^{-1}$ in the last century\footnote{Note that we are measuring time for an observer on Earth, we are not taking into account the $\sim 27000$ years the light travels to reach the Earth from \sgras.}.

\section{Conclusions}

In the present work, we expanded the previous X-ray studies of the Sgr A molecular complex by including all the XMM-Newton observations of the inner $\sim 35$ parsecs of the Galaxy, starting from 2000. To date, this is the most comprehensive description of the X-ray variability in the Sgr A complex, and it updates by adding at least 12 years of new data, previous similar works of \citet{Ponti2010,Capelli2012,Clavel2013,Terrier2018}. As of 2024, the Sgr A complex is still illuminated in the X-ray band, and variability was present in the whole 25-year monitoring period. Our analysis has shown that:
\begin{enumerate}
    \item As of 2024, the so-called Bridge is the brightest molecular cloud in the Sgr A complex; in particular, its eastern side (also known as M0.10-0.08 or the {\it Stick}). The Bridge is the only cloud inside the Sgr A complex in which the fluorescence signal has constantly increased during the last 25 years of monitoring. It was illuminated a few years later than the other clouds but shows an evolution on a comparable time scale.
    \item The superluminal propagation of the \Feka\ emission, first observed in the Bridge by \citet{Ponti2010}, is confirmed in the same region, in two different knots. The superluminal motion has propagated toward larger Galactic longitudes, as expected if the illuminating source is \sgras.
    \item The \Feka\ flux in all the other clouds has decreased since the start of the monitoring (G0.11-0.11), or has reached a peak and then start to faint (MC1 and MC2 and the Arches Cluster cloud). The \Feka\ intensity in the cloud associated with the position of the Arches Cluster has dropped by more than 70\% in the last decade. A residual emission is still present, and it cannot be excluded that is associated with cosmic rays bombardment. Similarly, in 2024, we found that the X-ray flux in MC2 has considerably dropped with respect to the past decades (80 \%), indicating that the wavefront of the past flare, which is illuminating the Sgr A complex, has almost passed these clouds.
    \item The 50 and 20 km/s molecular clouds show no fluorescence signal detection. This evidence is consistent with these two clouds being closer to \sgras\ and, therefore, already swept by any illuminating wavefront. Furthermore, the fact that the 50 km/s cloud is not illuminated provides a lower limit of the age of any energetic ($> 10^{36}$ erg s$^{-1}$) \sgras\ flare of $\sim 100$ years, given the geometry described in \citet{Coil2000}.
    \item Both a single and a multiple flares scenarios can describe the X-ray variability observed in the last 25 years. Previous shorter monitorings have pointed out that the main weakness of the single-flare scenario was that the densest clouds in the complex were not the brightest in the X-ray band. With 25 years of observations, this issue is completely overcome,  since the densest cloud (the Bridge), has reached an \Feka\ intensity superior to that observed in all other clouds. 
    \item Under the assumption that the clouds are illuminated by a single flare occurring some 200 years ago \citep[as suggested by the recent IXPE measurement,][]{Marin2023} and lasting 1.5 years, the Sgr A complex is located approximately 25 pc behind \sgras\ along the LOS. In this scenario, the Sgr A complex appears as a compact group of molecular clouds separated by a few parsecs from each other.
    \item The scenario in which two different flares illuminate the Sgr A complex can not be excluded. Since we found no evidence of cyclic or repeated fluorescence signal, the two flares should have been separated by at least 30 years. Moreover, the two flares must have emitted a similar amount of energy. Assuming the age of the flares suggested by \citet{Chuard2018}, the Bridge should be 10-15 pc closer to the Galactic center, separated, but still connected to the other clouds of the Sgr A complex, therefore forming a somehow peculiar long structure along the LOS. Additionally, in this scenario, the Sgr B2 complex should start to be illuminated by the second flare in a few decades ($\simeq 30$ years from now).
    \item Regardless of the number of flares that could illuminate the Sgr A complex, the molecular complex would not belong to the molecular ring encircling the CMZ (located at a radius of 100-200 pc from \sgras). We speculate that the Sgr A complex could be a streamer of molecular material on its way to the Galactic center from the CMZ.
    \item By using 2015 \chandra\ data, \citet{Churazov2017c} found that the PDF of the molecular hydrogen density shows an approximately log-normal distribution, with a variance $\sigma_s \simeq 0.7$. Our analysis confirms this result by taking into account the whole illuminated region in the past 25 years (adding, therefore, 15 more "slices"). The log-normal distribution of the density is, indeed, what is expected for a pure supersonic-driven flow. The PDF appears to be slightly skewed toward the highest densities.
\end{enumerate}

\begin{acknowledgements}
      We thank the anonymous referee for their comments that helped to improve the quality of the paper. We thank Michael C. H. Yeung for the helpful discussions. This work is based on observations obtained with XMM-Newton, an ESA science mission with instruments and contributions directly funded by ESA Member States and NASA. This project acknowledges financial support from the European Research Council (ERC) under the European Union’s Horizon 2020 research and innovation program HotMilk (grant agreement No. 865637), support from Bando per il Finanziamento della Ricerca Fondamentale 2022 dell’Istituto Nazionale di Astrofisica (INAF): GO Large program and from the Framework per l’Attrazione e il Rafforzamento delle Eccellenze (FARE) per la ricerca in Italia (R20L5S39T9). MCS acknowledges financial support from the European Research Council under the ERC Starting Grant ``GalFlow'' (grant 101116226) and from Fondazione Cariplo under the grant ERC attrattivit\`{a} n. 2023-3014. 
\end{acknowledgements}

\bibliographystyle{aa}
\bibliography{sample}

\end{document}